\documentclass[journal, twocolumn]{IEEEtran}
\usepackage{floatrow}
\usepackage[colorlinks,pdfstartview=FitH,citecolor=blue,linkcolor=blue,urlcolor=blue]{hyperref}

\usepackage{amsmath, amssymb, enumitem, bm, bbm, braket, tikz, tabulary, algorithm, listings, algorithmicx, amsthm, comment, balance}
\usepackage{subfigure} 

\usepackage[inline]{trackchanges}
\addeditor{swanand}
\addeditor{AS}
\addeditor{A1}
\addeditor{AN}

\usepackage{float}
\usepackage[noend]{algpseudocode}
\newcommand{\ie}{{\it i.e.}}
\newcommand{\eg}{{\it e.g.}}

\newcommand{\myapprox}{{\raise.17ex\hbox{$\scriptstyle\sim$}}}

\newcommand{\F}{\mathbb{F}} 

\newcommand{\E}[1]{\mathbb{E}\left[#1\right]}
\newcommand{\Prob}[1]{\text{P}\left\{#1\right\}}
\newcommand{\Exp}[1]{\mathrm{Exp}(#1)}

\newcommand{\Rli}[2]{\mathbb{R}^{#1}_{#2}} 

\newcommand{\dx}[1]{\textrm{d}#1} 

\makeatletter
\def\thm@space@setup{\thm@preskip=2pt
\thm@postskip=2pt}
\makeatother

\newtheorem{conjecture}{Conjecture}
\newtheorem{lemma}{Lemma}
\newtheorem{theorem}{Theorem}

\newtheorem{definition}{Definition}
\newtheorem{observation}{Observation}
\newtheorem{remark}{Remark}
\newtheorem{approximation}{Approximation}

\makeatletter
\renewenvironment{proof}[1][\proofname]{\par
  \pushQED{\qed}%
  \normalfont
  \topsep0pt \partopsep0pt 
  \trivlist
  \item[\hskip\labelsep
        \itshape
    #1\@addpunct{.}]\ignorespaces
}{%
  \popQED\endtrivlist\@endpefalse
  \addvspace{6pt plus 6pt} 
}
\makeatother

\allowdisplaybreaks[4]
\makeatletter
\newcounter{longaligned}
\newenvironment{longaligned}[1][]
 {%
  \stepcounter{longaligned}%
  \refstepcounter{equation}%
  \label{longaligned@\thelongaligned}%
  #1%
  \start@align\@ne\st@rredtrue\m@ne 
 }
 {\endalign}
\newcommand{\longalignedtag}{\tag{\ref{longaligned@\thelongaligned}}}
\makeatother

\makeatletter
\g@addto@macro\normalsize{%
  \setlength\abovedisplayskip{8pt}
  \setlength\belowdisplayskip{8pt}
  \setlength\abovedisplayshortskip{8pt}
  \setlength\belowdisplayshortskip{8pt}
}

\begin{document}

\title{Download Time Analysis for Distributed\\
Storage Codes with Locality and Availability}

\author{Mehmet Fatih Akta\c{s}, Swanand Kadhe, Emina Soljanin, and Alex Sprintson
\thanks{This paper was presented in part at ISIT'15 \cite{AnalyzingDownloadTimeOfAvailabilityCodes:KadheSS15} and Sigmetrics'17 \cite{SimplexQueues:AktasNS17}. 
M.~Akta\c{s} and E.~Soljanin are at Rutgers University, NJ 08854, USA, emails: {mfatihaktas@gmail.com, emina.soljanin@rutgers.edu},
S.~Kadhe is at UC Berkeley, CA 94720, USA, email: swnanand.kadhe@berkeley.edu, and
A.~Sprintson is at Texas A\&M University, TX 77843, USA. 
This material is based upon work supported by the NSF under Grant No.~CIF-1717314  and CIF-1718658. The work of Alex Sprintson (while serving at NSF) was supported by the NSF. Any opinions, findings, and conclusions or recommendations expressed in this material are those of the author(s) and do not necessarily reflect the views of the NSF.
}
}

\maketitle
\begin{abstract}

The paper presents techniques for analyzing the expected download time
in distributed  storage systems that employ systematic availability codes.
 These codes provide access to hot data through the systematic server containing the object and multiple recovery groups.
When a request for an object is received, it can be replicated (forked) to the systematic server and all recovery groups.
We first consider the low-traffic regime and present the close-form expression for the download time. By comparison across systems with availability, maximum distance separable (MDS), and replication codes, we demonstrate that availability codes can reduce download time in some settings but are not always optimal.
In the high-traffic regime, the system contains of multiple inter-dependent Fork-Join queues, making exact analysis intractable.
Accordingly, we present upper and lower bounds on the download time, and an M/G/1 queue approximation for several cases of interest.
Via extensive numerical simulations, we evaluate our bounds and demonstrate that the M/G/1 queue approximation has a high degree of accuracy.
\end{abstract}


\section{Introduction}
\label{sec:intro}
Distributed systems implement reliable storage despite failures by storing data objects redundantly across multiple servers. Replication is traditionally preferred for its simplicity. Maximum Distance Separable (MDS) codes are used when replication is costly.
Even though MDS codes maximize storage efficiency, they incur large communication overhead during object recovery.
This has motivated coding theorists to look for novel erasure codes that are {\it recovery-efficient}, see, e.g.,~\cite{Dimakis-Survey:11, Oggier:11, Gopalan:12}.
An important class of such codes is Locally Recoverable Codes (LRCs)~\cite{Oggier:11,Gopalan:12}. LRCs enable recovery of an object by accessing only a small group of servers, referred to as a {\it recovery group}. The number of servers in the largest recovery group is referred to as the code {\it locality}. 

A special class of LRCs, known as {\it availability codes}, have an additional property that each object has multiple, disjoint recovery groups~\cite{Tamo:14,Rawat:14,Pamies-Juarez:13,Wang-Zhang:14}.
Separate recovery groups allow simultaneous download of the same object by multiple users.
For instance, consider the binary Simplex code that encodes $\{f_1, f_2, f_3\}$ into $[f_1, ~f_2, ~f_3, ~f_1+f_2, ~f_1+f_3, ~f_2+f_3, ~f_1+f_2+f_3]$. 
This code is said to have {\it availability} three as each of $f_1$, $f_2$ and $f_3$ has three disjoint recovery groups. For example, $f_1$ can be recovered by reading both $f_2$ and $f_1+f_2$, or by reading both $f_3$ and $f_1+f_3$, or by reading $f_2+f_3$ and $f_1+f_2+f_3$. This code furthermore has {\it locality} two as each recovery group consists of at most two servers.

The notion of code availability was proposed 
with the goal of making the stored data more accessible (see, e.g.,~\cite{Rawat:14}).
Having multiple disjoint recovery groups for an object enables high data availability since each recovery group provides an additional way to retrieve the object. Thus, availability codes make it possible to assign multiple requests for the same object to different servers without blocking any request.
Consequently, availability codes have a significant potential to provide low-latency access for \emph{hot data}, i.e., objects that are frequently and simultaneously accessed by multiple users~\cite{Rawat:14}. 

It is important to quantify to what extent availability codes can reduce the download latency as compared to MDS and replication codes.
This is because accessing an object through one of its recovery groups requires downloading one object from each of the recovery servers.
A retrieval from a recovery group is therefore complete once the objects from {\it all} the servers in the recovery group are fetched. This means that the retrieval is slow even if the service is slow at only one of the servers.
In fact, service times in modern large-scale systems are known to exhibit significant variability \cite{Dremel:MelnikGL10, TailAtScale:DeanB13}, and thus, the download time from a recovery group can be significantly slower than that from a single server.
As an example, if service times at the servers are independent and exponentially distributed, mean time to download from a recovery group of size $r$ will scale (approximately) by $\ln r$ for large $r$.
Motivated by this practical challenge of service time variability, this paper presents techniques for analyzing the download latency of availability codes by leveraging tools from queuing theory.

\vspace{0.5ex}
\noindent
\textbf{Contributions and organization of the paper:}
We present techniques for analyzing the download time of individual data objects that are jointly encoded with a systematic availability code.
We assume the Fork-Join (FJ) access strategy, under which requests are {\it forked} upon arrival to the {\it systematic} server containing the requested object and all its recovery groups.
While other access schemes are possible, we focus our attention to the FJ strategy for the following reasons.
First, request service time at a systematic server is typically smaller than that at a recovery group. Thus, a user whose request is assigned to a recovery group would experience larger latency.
FJ strategy treats all the requests uniformly, resulting in a form of fairness.
Second, FJ strategy is widely adopted for download from coded storage systems (see, e.g.,~\cite{CodingForFastContentDownload:JoshiLS12, Codes&Qs:HuangPZ12,FJ:JoshiLS14,QueuesWithRed:JoshiSW15}).

We consider two arrival regimes:
(i) \emph{low-traffic regime} where each request is served before the new request arrives;
(ii) \emph{queuing regime} where requests can possibly overlap at the storage servers, which serve them through a first-in first-out queue.
First, for the low-traffic regime, we derive closed-form expressions for the distribution and expected value of download time for availability codes (Theorem~\ref{thm:T_avail_lowtraff}), along with replication (Lemma~\ref{lem:ET_rep}) and MDS codes (Lemma~\ref{lem:ET_MDS}). 
This enables us to compare availability codes with state-of-the-art erasure codes used in production systems, in particular the MDS code used in the Google file system and the locally recoverable code used in Windows Azure storage.
Next, for the queuing regime, we observe that the system consists of multiple inter-dependent Fork-Join queues. 
This results in the infamous state space explosion problem, which makes it intractable to perform an exact analysis.
Therefore, we establish upper and lower bounds on the download time.

Our key idea is to consider \textit{special cases} of our system by carefully imposing restrictions, and use download times of these restricted models to find the bounds. In particular, we consider the following two restricted models.
(1) Fork-Join Fixed Arrival (FJ-FA) model, in which every request arrival asks for the same object. 
This will hold for instance if the object popularities exhibit extreme skew, that is when the probability that a request asks for a particular object is one while it is zero for the rest of the objects.
(2)~Fork-Join Split-Merge (FJ-SM): in which requests are buffered upon arrival in a centralized First-come First-serve (FCFS) queue, and are fed to the system one by one only when all the servers are idle.
Leveraging these two models, we find lower and upper bounds on the performance of our Fork-Join system (Theorem~\ref{thm:T_FJGA_ulb} and Theorem~\ref{thm:ET_FJGA_lub}).
In addition, via numerical simulations, we compare the expected download time for availability codes with with various practical coding schemes.

In addition to the bounds, we also propose an M/G/1 queue approximation for FJ-FA systems with locality two. We then proceed by refining our approximation for systems with availability one.
In our analysis, we build on techniques from Markov processes and Renewal theory. Combination of the techniques we present is useful to study another type of Fork-Join system that is defined by an $(n, 2)$ MDS code \cite{MDSn2:AktasS18}.

\vspace{0.5ex}
\noindent
\textbf{Related work:}
\label{sec:rel_work}
Literature on download from coded storage has focused on downloading the \emph{complete set of data objects} that are jointly encoded with an MDS code. This is an important research question and has been studied extensively, see \eg~\cite{CodingForFastContentDownload:JoshiLS12, joshi2015queues, MeanFieldAnalysisCodingVsRep:LiRS16, LatencyAnalysisMDSorRep:Parag17, Rlog:JoshiSW17,FastCloud:LiangK14,Replication:JoshiSW15} and references therein.
This paper differs from this literature in two important aspects.
First, we are concerned with downloading individual objects. In practice, users are typically interested in only a subset of the stored data, namely \emph{hot data}.
This leads to skews in object popularities as shown by traces collected from production systems~\cite{CopingWithSkewedContentPopularityInMapreduce:AnanthanarayananAK11}.
Our model for data access incorporates the skewed object popularities and the notion of hot data.
Second, we focus on storage systems with availability codes. LRCs with availability have recently replaced MDS codes in production, \eg~\cite{ErasureCodingInAzureStorage:HuangSX12, XoringElephants:SathiamoorthyAP13}. It is important to understand the download performance when these new codes are at use.

The paper is organized as follows:
Section~\ref{sec:sys_model} explains the storage and data access model.
Section~\ref{sec:FJGA_lowtraff} analyzes the download time  under low-traffic regime. Section~\ref{sec:FJGA_hightraff} presents bounds on the download time under queuing regime.
Section~\ref{sec:FJ-variants} analyzes the FJ-FA and FJ-SM models to set up the background for the proofs of our bounds, and the proofs are presented in Section~\ref{sec:proof-main-theorems}.
Section~\ref{sec:mg1-approximations} presents an M/G/1 approximation for the FJ-FA system. In Section~\ref{subsec:mg1_r2} and~\ref{subsec:FJFA_r2_t1}, we consider the special cases with locality two and availability one, and obtain close approximations for the average download time for the FJ-FA model.
We skip some of the technical details in several proofs in Section~\ref{sec:mg1-approximations} for the sake of brevity. For the omitted details, we refer the reader to \cite{PhDThesis:Aktas20}.

\section{System Model}
\label{sec:sys_model}

\noindent 
\textbf{Data Storage Model}:
We consider a system that encodes $k$ equal sized objects $\F = \{f_1, \ldots, f_k\}$ into $n$ objects by an $(n,k)$ linear systematic erasure code, and then stores the $n$ encoded objects across $n$ servers. Each object $f_i$ is a symbol in some finite field $\mathbb{F}$.
The {\it storage overhead} of an $(n, k)$ linear code is defined as the inverse of the code rate, i.e., $n/k$.

We are interested in a special class of erasure codes for which each systematic server has $(r, t)$-{\it availability}~\cite{Tamo:14,Rawat:14,Pamies-Juarez:13,Wang-Zhang:14}.
A code is said to have $(r, t)$-availability if it ensures that failure of any systematic server can be recovered using one of the $t$ disjoint {\it recovery groups} of size $r$, where typically $r \ll k$.
We denote such an LRC as an $(n,k,r,t)$-LRC or an $(n,k,r,t)$ availability code.
We denote the $l$th recovery group for the $i$th server as $\Rli{l}{i}$, where $|\Rli{l}{i}| = r$.
The recovery group size $r$ is referred to as the code {\it locality}.
The $(r, t)$-availability allows retrieving an object $f_i$ in $t+1$ ways: either by downloading it from the systematic server or by downloading all the symbols in one of its $t$ recovery groups. 
In the example given in Section~\ref{sec:intro}, we used the $(n=7,k=3)$ binary Simplex code that encodes $\{f_1, f_2, f_3\}$ into $[f_1, ~f_2, ~f_3, ~f_1+f_2, ~f_1+f_3, ~f_2+f_3, ~f_1+f_2+f_3]$. This code has $(r=2, t=3)$-availability. Object $f_1$ can be accessed either at the systematic server or by recovering it from any of the three pairs of coded symbols $\{f_2, f_1+f_2\}$, $\{f_3, f_1+f_3\}$, $\{f_2+f_3, f_1+f_2+f_3\}$. Similarly for $f_2$ or $f_3$, there are one systematic server and three recovery groups.

\begin{figure}[t!]
  \includegraphics[width=0.8\textwidth]{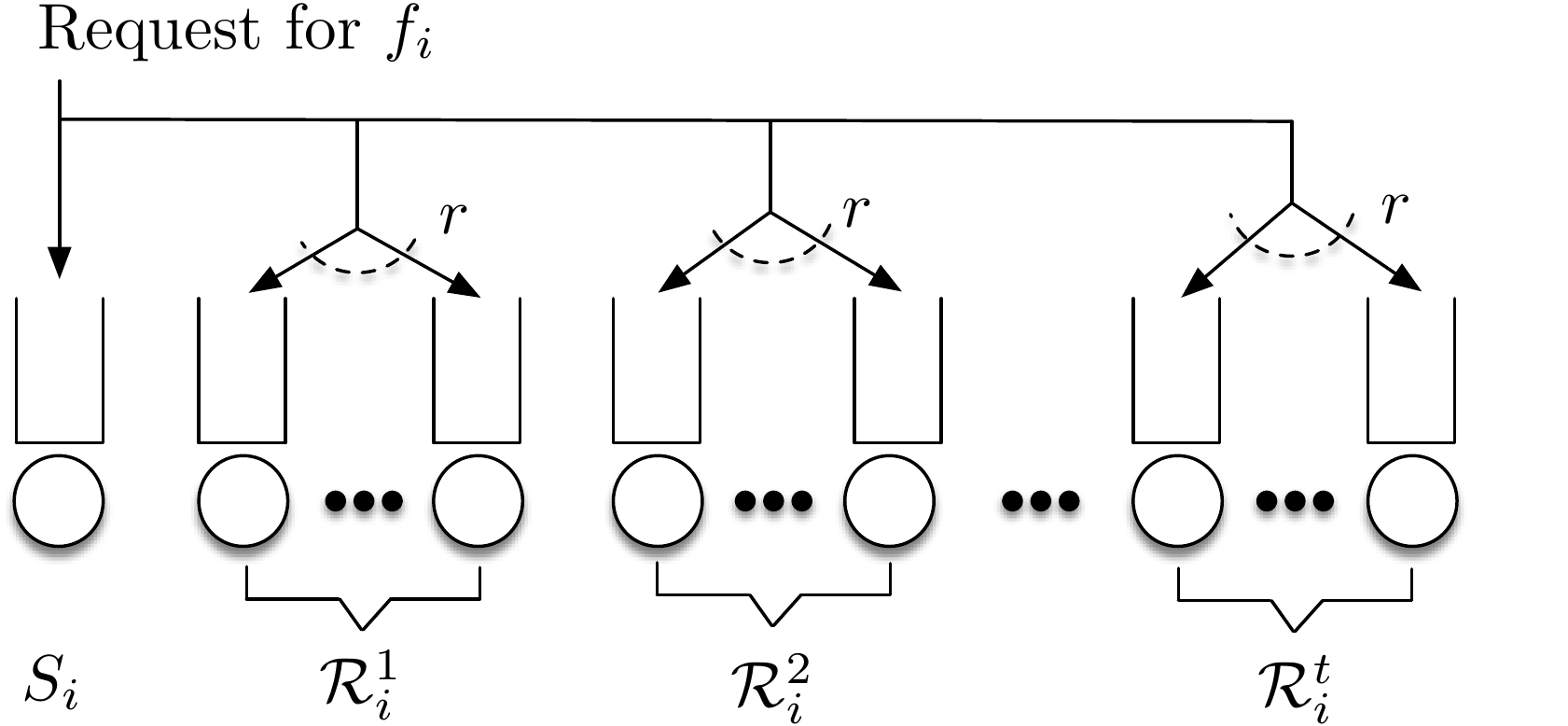}
  \caption{Fork-Join access model for $(n,k,r,t)$-LRC. Each request is split upon arrival into its systematic server and each of its $t$ recovery groups.}
\label{fig:FJ_availability}
\end{figure}

\vspace{0.5ex}
\noindent 
\textbf{Content Access Model}:
We consider the Fork-Join (FJ) model for data access.
FJ model for a system with an $(r, t)$-availability code consists of two levels of FJ queues.
In the first level, each request upon arrival is \emph{forked} (replicated) into $t+1$ copies, which are sent to the systematic server that stores the requested object and its $t$ recovery groups.
The request completes as soon as either its copy at the systematic server or a copy at any one of its recovery groups finishes service.
Once a request is completed, all its outstanding copies are immediately removed from the system.
The second level of FJ queues is formed at each recovery group.
A request copy that is assigned to a recovery group is forked into $r$ \textit{sub-copies} and each enters the queue at its respective server.
A request copy at a recovery group finishes when all its $r$ forked sub-copies finish service and \emph{join}.
(See Fig.~\ref{fig:FJ_availability} for an illustration.) 
We assume cancellation of the outstanding request copies (either in queue or in service), or reconstructing an object from the coded copies incurs negligible delay in comparison with the overall data access latency. Indeed, it is observed 
in practice that the latency of reconstructing an object is several orders of magnitude smaller than the overall data access latency~\cite{ErasureCodingInAzureStorage:HuangSX12}.

\vspace{0.5ex}
\noindent\textbf{Request Arrival and Service Models:}
For tractability, request arrival process is assumed to be Poisson with fixed rate $\lambda$.
We model object popularities as follows: each request arrival asks for object $f_i$ independently with probability $p_i$ for $i = 1, \dots, k$, where $0 \leq p_i \leq 1$ and $\sum_{i=1}^k p_i = 1$. 
We denote each request-$j$ as a pair $(t_j,o_j)$, where $t_j$ is the arrival time and $o_j$ is the requested object.
We consider two arrival regimes:\\
\ul{Low-traffic regime} in which the system completes serving a request before the next one arrives,  and can thus contain at most one request at any time. This case serves as a starting point for the download time analysis, and it is also useful to understand the download performance under low offered load.\\
\ul{Queuing regime}, in which subsequent requests might arrive before the current request(s) is finished.
In this case, system resources needs to be shared across the requests, and  multiple requests sent to a storage server are queued at the server.
We model resource sharing at the servers with FCFS queues.

Request copies are served at the servers by reading and streaming the requested object. We model this by a random \emph{service time}.
For tractability, service times are assumed to be independently and identically distributed (i.i.d.) across different requests and servers as Exponential random variables (r.v.'s) of rate $\mu$.
Note that, as argued in~\cite{FastCloud:LiangK14}, the shifted-exponential distribution is a good model of service times in data centers. 
However, the exponential distribution is analytically tractable because it is memoryless, and is commonly used to model service times in queuing theoretic analysis (see, \eg,~\cite{MeanFieldAnalysisCodingVsRep:LiRS16, LatencyAnalysisMDSorRep:Parag17,CodingForFastContentDownload:JoshiLS12,WhenDoesRedReduceLatency:ShahLR16}).
Insights derived from this study can serve as a stepping stone towards understanding the system performance under general service time distributions.


Our focus is on analyzing the  {\it download time} $T$ for an individual object, which essentially is the response time (or sojourn time) of a request. 
In particular, we define a request's download time as \mbox{$T = t_{\operatorname{departure}} - t_{\operatorname{arrival}}$,} where
$t_{\operatorname{arrival}}$ is the time the request arrives to system and
$t_{\operatorname{departure}}$ is the time the request download is completed.
Since we consider multiple coding schemes and variants of the FJ system, we typically denote the download time in system $x$ with r.v. $T_x$.

\vspace{0.5ex}
\noindent {\bf Notation:} $S$ denotes an Exponential random variable (r.v.) with rate $\mu$, i.e, $S\sim\Exp{\mu}$.
For r.v.'s $X$ and $Y$, $X \leq Y$ if and only if $\Prob{X > s} \leq \Prob{Y > s}$ for all $s$. Notation $X_{n:i}$ denotes the $i$th smallest of $n$ i.i.d. samples of a r.v. $X$. 
The beta function is given by $\beta(x,y)=\int_{0}^{1}v^{x-1}(1-v)^{y-1}\dx{v}$.

\section{Download Time Under Low-Traffic Regime}
\label{sec:FJGA_lowtraff}

\subsection{Download Time Characterization for Availability Codes}
\label{subsec:T_FJGA_lowtraff}
In the following, we characterize the download time $T_{\text{FJ-}(r, t)}$ in the FJ system implemented by an $(r,t)$-availability code.
\begin{theorem}
  Under the low-traffic regime,
  \begin{equation*}
  \begin{split}
    \Prob{T_{\text{FJ-}(r, t)} > s} &= \exp(-\mu s)\left(1 - (1 - \exp(-\mu s))^r\right)^t, \\
    \E{T_{\text{FJ-}(r, t)}} &= \frac{1}{\mu r} \beta(t+1, 1/r).
  \end{split}
  \end{equation*}
\label{thm:T_avail_lowtraff}
\end{theorem}
\begin{proof}
  $T_{\text{FJ-}(r, t)}=\min\{S, R_1, \ldots, R_t\}$ where $S$ is the service time at the systematic server and $R_i$ is the service time at the $i$th recovery group. 
  $R_i$ are independent and  distributed as $S_{r:r}$, since the request copy at a recovery group finishes service once all its copies at the $r$ recovery servers finish service.
  We have
  \begin{longaligned}[\label{eq:Pr_T_FJ_lowtraff}]
    \Prob{T_{\text{FJ-}(r, t)} > s} &= \Prob{\min\{S, R_1, \ldots, R_t\} > s} \\
    &\stackrel{(a)}{=} \Prob{S > s} \left(1 - \Prob{S \leq s}^r\right)^t     \longalignedtag\\
    &\stackrel{(b)}{=} \exp(-\mu s)\left(1 - (1 - \exp(-\mu s))^r\right)^t
  \end{longaligned}
  where (a) follows from the independence of the service times and (b) comes from substituting $\Prob{S > s} = \exp(-\mu s)$.
  
  Observing that $T_{\text{FJ-}(r, t)}$ is a non-negative r.v., we obtain
  \begin{longaligned}[\label{eq:ET_FJ}]
    \E{T_{\text{FJ-}(r, t)}} &= \int_{0}^{\infty} \Prob{T_{\text{FJ-}(r, t)} > s} \dx{s}\longalignedtag \\
    &\stackrel{(a)}{=} \frac{1}{\mu r} \int_{0}^{1} v^{t}(1-v)^{\left(1/r-1\right)} \dx{v}
    \stackrel{(b)}{=} \frac{1}{\mu  r} \beta\Bigl(t+1, \frac{1}{r}\Bigr)
  \end{longaligned}
  where (a) comes from \eqref{eq:Pr_T_FJ_lowtraff} and $1 - (1 - \exp(-\mu s))^r = v$,
  and (b) follows from the definition of the beta function.
\end{proof}

\begin{figure}[t!]
  \includegraphics[width=0.75\textwidth]{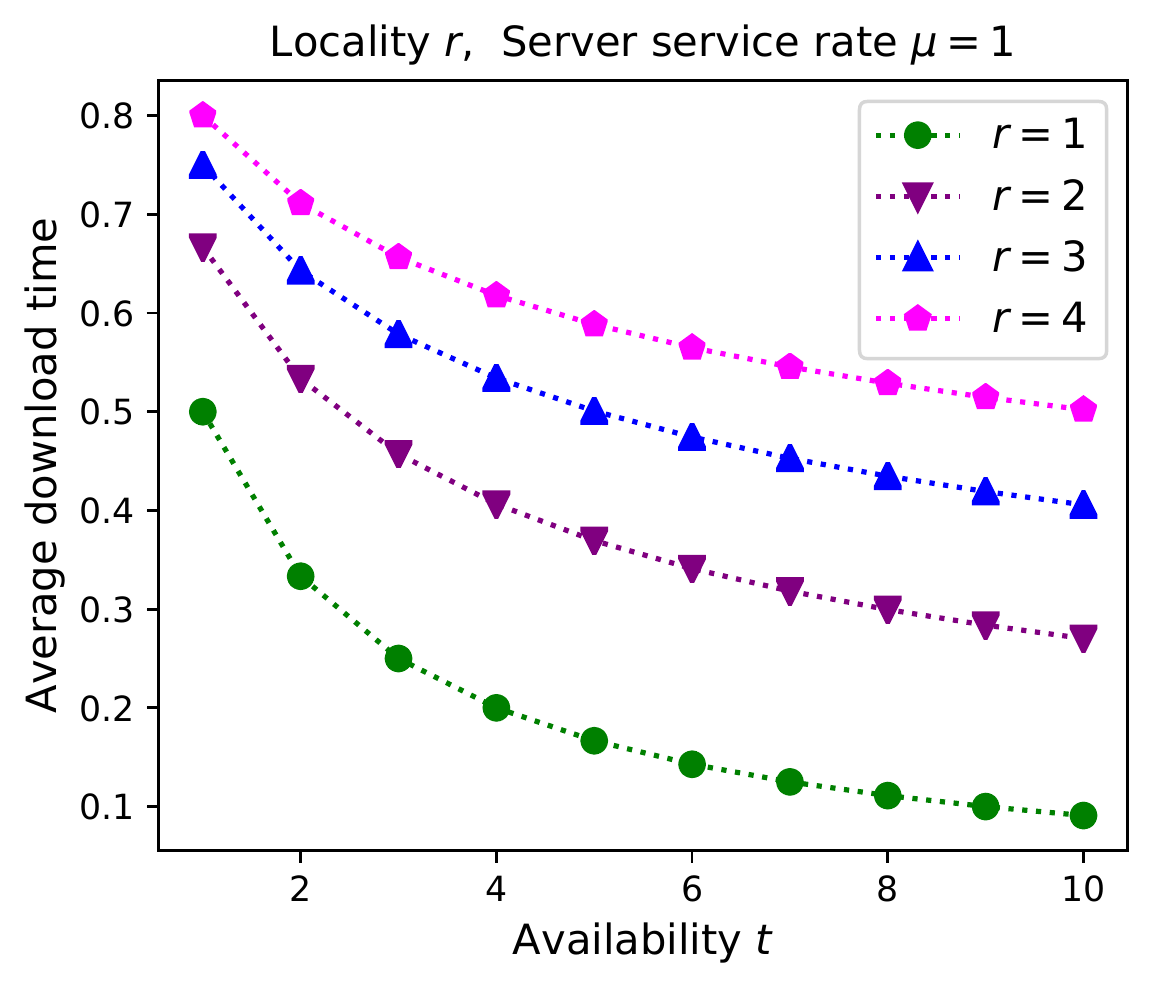}
  \caption{Average download time under the low-traffic regime with respect to code locality $r$ and availability $t$.
  }
\label{fig:ETlowtraff_wrt_r_t}
\vspace{-2mm}
\end{figure}

\begin{remark}
  Expression for $\E{T_{\text{FJ-}(r, t)}}$ allows us to examine the effect  $t$ on the average download time while fixing $r$.
  Using the equality $\beta(x, y+1) = \frac{y}{x+y}\beta(x,y)$, it is straightforward to verify that the relative reduction in average download time per increment in $t$ is given by $(r(t+1) + 1)^{-1}$, which is approximately $1/rt$ for large $t$ and $r$.
  Thus incrementing $t$ yields diminishing returns in reducing the average download time. This can be observed in Fig.~\ref{fig:ETlowtraff_wrt_r_t}, which plots the average download time with respect to $r$ and $t$.
  Observe that increasing the locality $r$ significantly slows down the download.
\label{rem:dimingreturn_in_ET_by_incing_t}
\end{remark}

\noindent{\bf Comparison with Replication Codes:} 
Replication schemes are an important contestant of availability codes as they are commonly deployed in production, see \eg,~\cite{Dynamo:DecandiaHJ07}.
We here consider a $t_{\operatorname{rep}}$-replication code that stores $t_{\operatorname{rep}}$ copies for each of the $k$ objects. 
For a fair comparison, we assume the cumulative service rate (sum of the service rates across all the servers) is fixed.
The proof of the following lemma follows from the order statistics, similar to the proof of Theorem~\ref{thm:T_avail_lowtraff}.
\begin{lemma} 
  In a system with $t_{\operatorname{rep}}$-replication code and service times i.i.d. as $\Exp{\mu}$, the average download time under the low-traffic regime is $\E{T_{\text{FJ-}t_{\operatorname{rep}}}} = 1/(t_{\operatorname{rep}} \mu)$.
  If the cumulative service rate of the system is the same as that of a system with an $(n, k, r, t)$-LRC, then $\E{T_{\text{FJ-}t_{\operatorname{rep}}}} = k/(n \mu)$.
\label{lem:ET_rep}
\end{lemma}

\begin{remark}
  When the cumulative service rate is equal to that of an $(n,k)$ code, the average download time in the $t_{\operatorname{rep}}$-replication system depends only on $k/n$, and not on $t_{\operatorname{rep}}$.
  \label{rem:availability_vs_replication}
\end{remark}

\noindent {\bf Comparison with MDS Codes:}
An $(n, k)$ code is said to be MDS if any $k$ of the $n$ coded objects are sufficient to reconstruct the original $k$ objects.
Recent work have shown that MDS codes are faster than replication in downloading the entire set of $k$ objects, see \eg,~\cite{CodingForFastContentDownload:JoshiLS12}. It is therefore natural to evaluate MDS codes in downloading the individual objects.

In the FJ access model for MDS coded storage, each request is replicated upon arrival into $n$ copies, which are then assigned across all the $n$ servers.
A request is completed once either its copy at the systematic server finishes service or any $k$ out of the remaining $n-1$ recovery servers jointly finish serving the request.
The download time $T_{\text{FJ-}(n, k)}$ is given by $\min\{S, ~S_{(n-1):k}\}$.
This allows us to derive the average download time under the low-traffic regime as follows. 
The proof follows from the order statistics, similar to Theorem~\ref{thm:T_avail_lowtraff}.

\begin{lemma} 
  In a system with an $(n, k)$-MDS code and service times i.i.d. as $\Exp{\mu}$, the expected download time under the low-traffic regime is $\E{T_{\text{FJ-}(n, k)}} = {k}/{(n \mu)}$.
\label{lem:ET_MDS}
\end{lemma}

\begin{remark}
  The average download time in an $(n,k)$-MDS coded system depends only on $k/n$, and is the same as that in a $t_{\operatorname{rep}}$-replication coded system when the cumulative mean service rates of both the systems are the same.
\label{rem:comparison-MDS}
\end{remark}


\begin{table}
  \floatbox[{\capbeside\thisfloatsetup{capposition={top},capbesidewidth=\textwidth}}]{table}
  \caption{{\sc Comparison of Three Commercial Codes and an Availability Code Based on Three Performance Metrics.\label{tbl:comparison-fountain}}}
\vspace{2ex}
\begin{tabular}{| c | c | c | c |}
\hline
Erasure Code & $\E{T} \mu$ & \begin{tabular}{@{}c@{}}Storage \\  Overhead\end{tabular} & \begin{tabular}{@{}c@{}}Fault \\  Tolerance\end{tabular} \\ 
\hline
$3$-replication & $0.33$ $(0.67)$ & 
$3\times$ &  $2$ \\
\hline
$(9,6)$-MDS code & $0.67$ $(0.67)$ & 
$1.5\times$ & $3$ \\
\hline
$(10,6,3,1)$-LRC & $0.6$ $(0.83)$ & 
$1.5\times$ & $3$ \\ 
\hline
$(14,6,2,3)$-LRC & $0.45$ $(0.71)$ & 
$2.33\times$ &  $3$ \\
\hline
\end{tabular}
\\[2ex]
        \footnotesize
       The average download time is normalized with respect to the service rate $\mu$.
%
\end{table}

\subsection{Performance Comparison of Erasure Codes}
\label{sec:holistic-comparison}
We here compare availability codes with the following state-of-the-art erasure codes:
(i) 3-replication, which is commonly used in many distributed storage systems, \eg, Amazon's Dynamo \cite{Dynamo:DecandiaHJ07};
(ii) $(9,6)$-MDS code, which is used in Google file system~\cite{AvailabilityGloballyDistributedStorage:Ford:10};
(iii) $(10,6,2,1)$-LRC, which is used in Windows Azure Storage~\cite{ErasureCodingInAzureStorage:HuangSX12}; and 
(iv) $(14,6,2,3)$-availability code obtained as a direct sum of two $(7,3,2,3)$-Simplex codes. Binary Simplex code, which is a {\it dual} of the Hamming code, forms a well-known class of availability codes with locality two~\cite{BoundsOnSizeOfLRCs:CadambeM15}, and a direct sum of simplex codes yields codes with availability~\cite{RateOptimalityOfSimplex:KadheC17}.
We here adopt the same data access and request service model introduced in Section \ref{sec:sys_model}. This means requests are replicated to all servers upon arrival, and the request service times are i.i.d. as $\Exp{\mu}$.
We use Lemma~\ref{lem:ET_MDS} and Theorem~\ref{thm:T_avail_lowtraff} to compute the average download time for the systems with the erasure codes listed above.

In order to get a holistic performance comparison, we consider the following metrics in addition to average download time: (i) storage overhead and (ii) fault tolerance. In particular, it is well-known that a coding scheme that encodes $k$ objects across $n$ nodes with minimum distance $d$ can tolerate any $d-1$ node failures. We refer to $d-1$ as the fault tolerance of the coded system.\footnote{More generally, reliability analysis can be done in terms of mean time to data loss (MTTDL) using standard techniques, see e.g.,~\cite{AvailabilityGloballyDistributedStorage:Ford:10,ErasureCodingInAzureStorage:HuangSX12,XoringElephants:SathiamoorthyAP13}.
It is well-known that MTTDL grows exponentially with the minimum distance. For simplicity, we focus on the fault-tolerance instead of MTTDL.}
We emphasize that, even though the storage overhead and fault tolerance have been known to be important metrics in evaluating real-world distributed storage systems (see, \eg,~\cite{AvailabilityGloballyDistributedStorage:Ford:10,ErasureCodingInAzureStorage:HuangSX12,XoringElephants:SathiamoorthyAP13}), the critical metric of (average) download time is missing in the literature.

We present a detailed comparison in Table~\ref{tbl:comparison-fountain}. The $(14,6,2,3)$-LRC with $(2,3)$ availability achieves smaller average download time as compared to the state-of-the-art erasure codes, at the expense of worse storage overhead. On the other hand, its storage overhead is smaller compared to $3$-replication, at the cost of slightly worse download time. Indeed, it achieves a favorable trade-off between the storage overhead and the average download time, with large fault-tolerance.
Availability codes are therefore attractive for storing hot data that requires small download latency. 
On the other hand, when the cumulative service rate in the system is kept fixed, $(9,6)$-MDS code achieves the smallest download time as well as the smallest storage overhead, with high fault-tolerance. Hence, for storage systems with limited cumulative service rate, MDS codes are favorable candidates.

\section{Download Time Under Queuing Regime}
\label{sec:FJGA_hightraff}
In this section we remove the assumption that at most one request is present in the system at any time.
In this case, multiple requests might be simultaneously present in the system, hence queues might build up at the servers.
In order to highlight the structure of the queues in our system, let us first review the notion of an $(\ell,m)$-Fork-Join (FJ) queue~\cite{CodingForFastContentDownload:JoshiLS12}.
In an $(\ell, m)$-FJ queue, each request is forked into $\ell$ copies upon arrival, and a request completes once any $m$ out of its $\ell$ copies finish service. Once a request completes, its remaining $\ell-m$ copies are immediately removed from the system. 

In the system that uses availability codes, there are two levels of FJ queues present under the FJ access model.
In the {\it outer} level, each arriving request is replicated into $t+1$ copies, which are then sent to the systematic server and its $t$ recovery groups. A request completes as soon as any one of its copies finishes service, hence forming a $(t+1, 1)$-FJ queue.
In the {\it inner} level, request copies sent to a recovery group fork into $r$ sub-copies and complete once all the forked copies finish service. Thus, each recovery group acts as an $(r, r)$-FJ queue. Note that FJ queues at the recovery groups are not independent due to the cancellation of outstanding request copies.

Analysis of the FJ queues is a long standing open problem. Its average response time is known only for the simplest case of $(2, 2)$-FJ queue \cite{FJTwoservers:FlattoH84, Nelson-Tantawi}.
Moreover, the inner FJ queues implemented by the recovery groups in our system are inter-dependent.
This inter-dependence makes it intractable to exactly analyse our system.
Our approach to understand the system performance is therefore deriving bounds and approximations on the download time.

\subsection{Bounds on Download Time}
\label{subsec:ub_w_SM_FA}
We here present bounds on the download time by considering more restricted counterparts of our system. These restrictions are carefully imposed on the system to make it possible to find exact or approximate expressions.

\noindent{\bf Fork-Join General Arrival (FJ-GA):} 
We refer to our availability coded system under the Fork-Join access model and object popularity probabilities $(p_1, \dots, p_k)$ as the Fork-Join General-Arrival (FJ-GA) system.

\noindent{\bf Fork-Join Fixed Arrival (FJ-FA):} 
In this restrictive model, every request arrival within a busy period asks for the same object. A busy period refers to the time interval during which there is at least one request in the system.
This will hold for instance if the object popularities exhibit extreme skew, i.e.,
when requests ask for a particular object with probability one.
In general, the FJ-FA model is useful to understand the performance of the FJ model under highly skewed object popularities, which is typical in practice \cite{CopingWithSkewedContentPopularityInMapreduce:AnanthanarayananAK11}.

\noindent{\bf Fork-Join Split-Merge (FJ-SM):} 
In this restrictive model, requests are buffered in a centralized FCFS queue, and are sent to service one by one only when all the servers are idle.

Recall that each download request-$i$ is of the form $(t_i,o_i)$ where $t_i$ is the arrival time and $o_i$ is the requested object. The download time refers to the steady state system response time, i.e., the time spent by an individual request in the system.
In the following, we denote the download time in system $x$ under an aggregate arrival rate of $\lambda$ as $T_{x, \lambda}$. We avoid writing $\lambda$ explicitly when the arrival rate is clear from the context.

\begin{theorem}
  Download time distributions for the FJ-GA, FJ-FA, FJ-SM systems satisfy the following inequalities:
  \begin{equation}
  \begin{split}
    \Prob{T_{\text{FJ-GA}, \lambda} > t} &\stackrel{(a)}{\geq} \sum_{i=1}^{k} p_i \;\Prob{T_{\text{FJ-FA}, \lambda p_i} > t} \\
    &\stackrel{(b)}{\geq} \sum_{i=1}^{k} p_i \;\exp\left(-\left((t+1)\mu - p_i \lambda\right)t\right),
  \end{split}
  \label{eq:T_FJGA_lb}
  \end{equation}
  \vspace{-12pt}
  \begin{equation}
  \begin{split}
    \Prob{T_{\text{FJ-GA}, \lambda} > t} &\stackrel{(c)}{\leq} \Prob{T_{\text{FJ-SM}, \lambda} > t}, \\
    \Prob{T_{\text{FJ-FA}, \lambda} > t} &\stackrel{(d)}{\leq} \Prob{T_{\text{FJ-SM}, \lambda} > t}.
  \end{split}
  \label{eq:T_FJGA_ub}
  \end{equation}
\label{thm:T_FJGA_ulb}
\end{theorem}

\begin{lemma}
  When code locality is two, $\Prob{T_{\text{FJ-GA}, \lambda} > t} \leq \Prob{T_{\text{FJ-FA}, \lambda} > t}$ for all $\lambda$ and $t$.
\label{lm:T_FJGA_leq_T_FJFA}
\end{lemma}

\begin{remark}
  It is possible to interpret the lower bound $(a)$ as the response time in a system of $k$ separate FJ-FA's, which are fed by splitting the Poisson arrival process across with the probabilities $p_1, \dots, p_k$. In fact, this lower bound is derived using this interpretation as outlined later in Section~\ref{sec:proof-main-theorems}.
  In this system of $k$ FJ-FA's, minimum response time is achieved when the load is perfectly balanced across the FJ-FA's, that is when $p_i = 1/k$ for $i = 1, \dots, k$. The response time will increase $(p_1, \dots, p_k)$ becomes skewed.
  This means the lower bound $(a)$ will take its minimum possible value when $p_i = 1/k$ for $i = 1, \dots, k$. The value of the lower bound will increase as $(p_1, \dots, p_k)$ becomes skewed, e.g., when $p_1 > p_i$ for $i = 2, \dots, k$.
  Lemma~\ref{lm:T_FJGA_leq_T_FJFA} together with the upper bounds $(c)$ and $(d)$ says that, when the code locality is two, the lower bound $(a)$ becomes equal to the upper bound $(c)$ when the object popularities are polarized, i.e., when every arrival asks for the same object.
  Then, when code locality is two, the upper bound $(c)$ suggests that the performance of FJ-GA becomes worst when the object popularities are highly skewed.
  The bounds in Theorem~\ref{thm:T_FJGA_ulb} become looser as the request arrival rate gets larger. Figure~\ref{fig:ET_FJFA_r2_t1_hightraff_approx} shows this for the system with availability one and locality two. 
  For this reason, we also find an approximation for the download time in Section~\ref{sec:mg1-approximations}.
\label{rm:FJFA_mg1}
\end{remark}

Theorem~\ref{thm:T_FJGA_ulb} gives us the following bounds on the average download time in the FJ-GA system.
We note that the bounds on $\lambda$ in the theorem statement are due to stability constraints. 
\begin{theorem}
The average download time in the FJ-GA system $\E{T_{\text{FJ-GA}}}$ is bounded as follows.
  \begin{longaligned}
   \label{eq:ET_FJGA_lub}
    \E{T_{\text{FJ-GA}}} &\geq \sum_{i=1}^{k} \frac{p_i}{(t+1)\mu - p_i \lambda},\longalignedtag \\
    \E{T_{\text{FJ-GA}}} &\leq \eta
    + \frac{\lambda \sum_{j=0}^{t}\binom{t}{j}(-1)^j \sum_{l=0}^{r j}(-1)^l\binom{r j}{l} (l+1)^{-2}}{\mu^2 (1 - \lambda \eta)},
  \end{longaligned} 
  where $\eta = \frac{1}{\mu r} \beta\Bigl(t+1, \frac{1}{r}\Bigr)$.
  The lower bound holds for $\lambda < (t+1)\mu/\max\{p_1, \dots, p_k\}$, and the upper bound, for $\lambda < 1/\eta$.
\label{thm:ET_FJGA_lub}
\end{theorem}
We defer proving Theorem~\ref{thm:T_FJGA_ulb} and Theorem~\ref{thm:ET_FJGA_lub} 
to Section~\ref{sec:proof-main-theorems} after we establish the characteristics of the FJ-FA and FJ-SM systems in Section~\ref{sec:FJFA} and \ref{sec:FJSM}, which are used in the proofs.



\subsection{Simulation Results}
\label{subsec:simulations}
\begin{figure}[ht]
  \centering
  \subfigure{\includegraphics[width=0.8\textwidth]{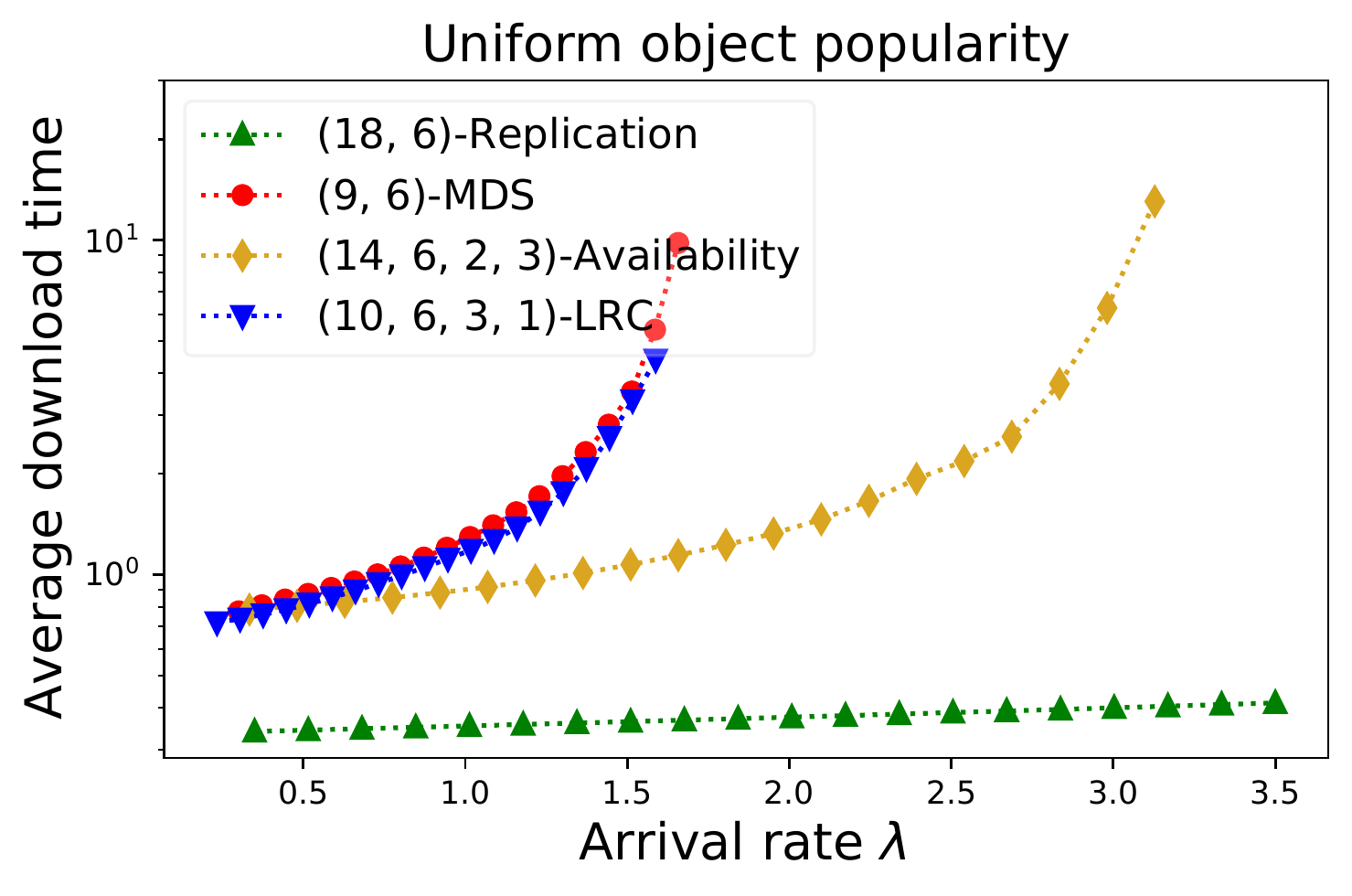}}
  \subfigure{\includegraphics[width=0.8\textwidth]{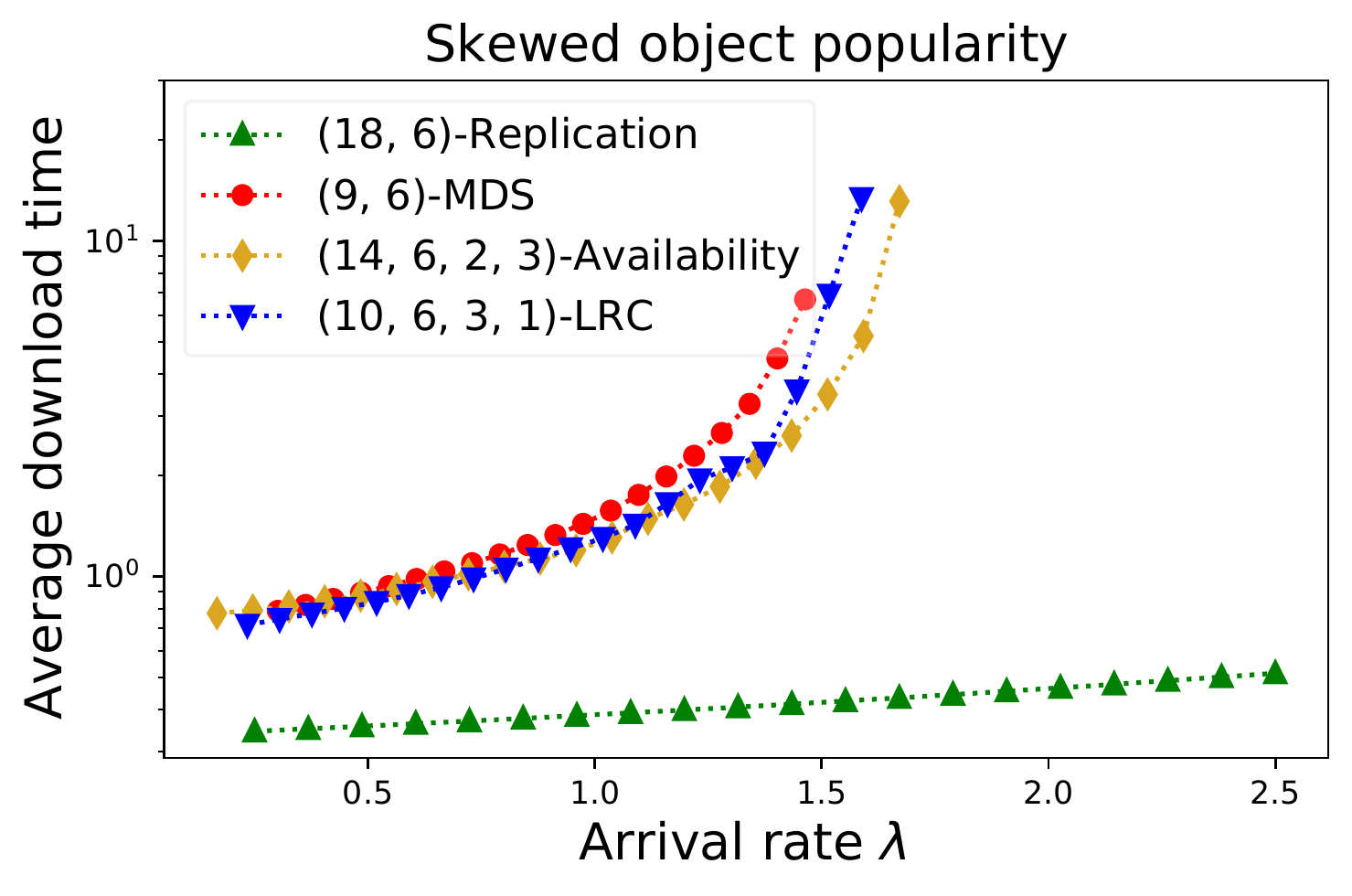}}
  \caption{Average download time versus $\lambda$ for the coding schemes in Table~\ref{tbl:comparison-fountain}. 
  At the top, all objects have equal popularity.
  At the bottom, $1/3$ of the objects have $90\%$ of the popularity and the rest share $10\%$ of popularity equally, e.g., $p_1 = p_2 = 0.45$, and $p_3 = p_4 = p_5 = 0.05$ for $(14, 6, 2, 3)$-availability.
  }
\label{fig:sim}
\vspace{-4mm}
\end{figure}

Fig.~\ref{fig:sim} shows the average download time vs. request arrival rate for 
the coding schemes given in Table~\ref{tbl:comparison-fountain}.
Note that all the four coding schemes used here store $6$ objects. However the number of servers $n$ dictated by the code varies across the schemes. 
For a fair comparison, we also plot results when  the cumulative service rate is fixed at 10, which is evenly allocated across the servers. 
For instance, server service rate is $1$ in the system with $(10, 6, 3, 1)$-LRC while it is $10/14$ in the system with $(14, 6, 2, 3)$-Availability code.
For high arrival rates, these systems are ordered in terms of their average download time as 
$(18, 6)$-Replication $<$ $(14,6,2,3)$-Availability $<$ $(10,6,3,1)$-LRC $<$ $(9,6)$-MDS.
Notice that this is the same ordering given by their storage efficiency.
This means availability codes in the Fork-Join access model serve as an intermediate point in the storage vs. download time tradeoff between the two extremes: MDS and replication codes.
This observation holds regardless of the skews in object popularities.

\section{Variants of the Fork-Join System}
\label{sec:FJ-variants}
\subsection{Fork-Join System in the Fixed-Arrival Model (FJ-FA)}
\label{sec:FJFA}
Recall that, in the FJ-FA model, all request arrivals within a busy period ask for the same object.
Therefore, each server in the system acts either as the systematic server or as a recovery server within the same busy period. Further, only one server takes the role of systematic server in the busy period. 

\noindent\textbf{Leading and Slow Servers:} A request copy is forked into $r$ sibling servers once it arrives to a recovery group. Some of the servers within the recovery group can be ahead of their siblings in service at any given time. We refer to such servers as \emph{leading}, and to those behind with service as \emph{slow} servers. For instance, in the example shown in Fig.~\ref{fig:FJFA_r2_t1_sys_snapshot} (in Section~\ref{subsec:FJFA_r2_t1}), the server that hosts $a+b$ is ahead of its sibling (the one that hosts $b$) by one request copy. 
In this case,  we say the server hosting $a+b$ is leading while the server hosting $b$ is slow.

Observe that up to $1+r(t-1)$ requests can be served simultaneously, and different copies of the same request can start service at different times.
These properties significantly complicate the system analysis. In order to address this challenge, we redefine the request service start times as follows.
\begin{definition}
  We say  that a {\it request is at the head of the line (HoL)} once all its copies remaining in the system are in service, and 
  the {\it request starts service} once it moves to HoL.
\label{def:FJFA_reqservstart}
\end{definition}

Under this definition, we have the following
observation. 
\begin{observation}
  In the FJ-FA system, requests depart the system in the order they arrive and there can be at most one request in service at any time.
\label{obv:FJFA_dep_order}
\end{observation}

The first part follows from the fact that request copies depart in the order of arrival both at the systematic server and at any recovery group.
To see the second part, note that, for two requests to be simultaneously in service, all remaining copies of each must be in service simultaneously, which is impossible given that requests depart
in the first-in first-out order.


At each recovery group, a request can have up to $r-1$ of its copies depart before the request moves to HoL. We refer to these  as \emph{early departing copies}.
If $d$ copies of a request depart early at a recovery group, service time of the request copy at that recovery group will be distributed as $S_{r-d:r-d}$.
More generally, let $d_i$ denote, for a request, the number of its early departing copies at the $i$th recovery group.
Then the service time for the request, once it moves to HoL, is given as
$S_{d_1, \dots, d_t} = \min\left\{S, ~S_{r-d_1:r-d_1}, ~\dots, ~S_{r-d_t:r-d_t}\right\}$.
For a multiset $\{d_1, \dots, d_t\}$ and each $d_i \in \{0, 1, \dots, r-1\}$, let $\nu_d$ denote the number of occurrences of $d$ in the multiset.
Given that the recovery groups are indistinguishable, $S_{d_1, \dots, d_t}$ is solely determined by the vector $\bm{\nu} = (\nu_0, \nu_1, \dots, \nu_{r-1})$, and we say in this case that the request has \emph{type-$\bm{\nu}$} service time, and denote it as $S_{\bm{\nu}}$. The set of all possible $\bm{\nu}$ is given by
\begin{equation}
  \mathbb{N} = \Bigl\{(\nu_0, \nu_1, \dots, \nu_{r-1}) \mid \sum_{0}^{r-1} \nu_i = t, ~\nu_i \in \mathbb{Z}_{\geq 0}, \Bigr\}
\label{eq:set_nu}
\end{equation}
where $\mathbb{Z}_{\geq 0}$ is the set of non-negative integers.
We formalize these observations in the following lemma.
\begin{lemma}
  In the FJ-FA system, the set of all possible request service time distributions is
  $\mathbb{S} = \left\{S_{\bm{\nu}} \mid ~\bm{\nu} \in \mathbb{N}\right\}$,
  where $\mathbb{N}$ is given in \eqref{eq:set_nu} and $|\mathbb{S}| = |\mathbb{N}| = \binom{t+r-1}{r-1}$.
  The distribution function for the type-$\bm{\nu}$ service time is 
  \begin{equation*}
    \Prob{S_{\bm{\nu}} > s} = \exp(-\mu s) \prod_{d=0}^{r-1} \left(1 - (1 - \exp(-\mu s))^{r-d}\right)^{\nu_d}.
  \end{equation*}
\label{lm:Stype_FJFA}
\end{lemma}

Observe from the expression of $\Prob{S_{\bm{\nu}} > s}$ that the more copies of a request depart early (before the request moves to HoL), the faster its service will be.
Recall that $\sum_i \nu_i = t$ for any $\bm{\nu} \in \mathbb{N}$. Distribution $\Prob{S_{\bm{\nu}} > s}$ gets smaller as the mass $t$ of $\bm{\nu}$ is shifted on $\nu_i$ with larger $i$. Thus, we have the following partial ordering between the possible request service time distributions: $S_{\bm{\nu}} > S_{\bm{\nu}^\prime}$,
if there exists a $j$ such that $\nu_i \leq \nu_i^\prime$ for all $i \geq j$.
This ordering implies that $S_{\bm{\nu}}$ is the \textit{fastest} and the \textit{slowest} when $\bm{\nu}$ is equal to $(0, \dots, 0, t)$ and $(t, 0, \dots, 0)$ respectively.
We refer to $S_{(0, \dots, 0, t)}$ as $S_{\text{fastest}}$, and $S_{(t, 0, \dots, 0)}$ as $S_{\text{slowest}}$. Their distributions are
\begin{equation}
\begin{split}
  \Prob{S_{\text{fastest}} > s} &= e^{-\mu (t+1) s}, \\
  \Prob{S_{\text{slowest}} > s} &= e^{-\mu s} (1 - (1 - e^{-\mu s})^r)^t.
\end{split}
\label{eq:S_fastest_slowest}
\end{equation}

\subsection{Fork-Join System in the Split-Merge Model (FJ-SM)}
\label{sec:FJSM}
Recall that in the FJ-SM system, requests wait in a centralized FCFS queue and are admitted to service one at a time.
FJ-SM therefore does not introduce any dependence between the service time distributions of different requests.
In fact, it is straightforward to see that
the FJ-SM system implements an M/G/1 queue with an arrival rate of $\lambda$ and the service time distribution $T_{\text{FJ-}(r, t)}$ given in Theorem~\ref{thm:T_avail_lowtraff}. 

In FJ-SM, all servers are blocked until the request at the HoL finishes service, whereas in FJ-GA or FJ-FA, each server can independently start serving from their queues.
FJ-SM therefore performs slower than both FJ-GA and FJ-FA.
This serves us to find an upper bound on the download time in FJ-GA and FJ-FA as stated below.
It should be noted that the Split-Merge model is also used in~\cite{CodingForFastContentDownload:JoshiLS12} for deriving upper bounds on the response time of $(n, k)$-FJ queue.
\begin{lemma}
The download time in FJ-SM is stochastically dominant over that in FJ-FA and FJ-GA, i.e.,
$\Prob{T_{\text{FJ-GA}} > t} \leq \Prob{T_{\text{FJ-SM}} > t}$ and  $\Prob{T_{\text{FJ-FA}} > t} \leq \Prob{T_{\text{FJ-SM}} > t}$.
\label{lm:T_FJ_ub_wSM}
\end{lemma}

\noindent
\textbf{Fast-Split-Merge Model}:
Observe that the request service time distribution in FJ-SM, given in Theorem~\ref{thm:T_avail_lowtraff}, 
is the same as $\Prob{S_{\text{slowest}} > s}$ in \eqref{eq:S_fastest_slowest}.
Recall that $S_{\text{slowest}}$ denotes the slowest possible request service time distribution in the FJ-FA system.
In other words, FJ-SM is a modification of FJ-FA such that it forces every request to have the slowest possible service time.

We next consider the \emph{Fast-Split-Merge model}, denoted as FJ-FSM, which
operates at the other extreme of FJ-SM and serves every request with the fastest possible service time $S_{\text{fastest}}$ in \eqref{eq:S_fastest_slowest}.
Using Observation~\ref{obv:FJFA_dep_order}, it is straightforward to show that the FJ-FSM model implements an M/M/1 queue with an arrival rate of $\lambda$ and service time distribution $\Exp{(t+1)\mu}$. 
Just as we used FJ-SM to find an upper bound on $T_{\text{FJ-FA}}$ (Lemma~\ref{lm:T_FJ_ub_wSM}), we next use FJ-FSM to find a lower bound.
\begin{lemma}
  $\Prob{T_{\text{FJ-FA}} > x} \geq \exp\left(-\left((t+1)\mu - \lambda\right)x\right)$ for $\lambda < (t+1)\mu$.
\label{lm:T_FJFA_lb_wFSM}
\end{lemma}
\begin{proof}
  FJ-FA implements a FCFS queue (Observation~\ref{obv:FJFA_dep_order}), and each request is served with one of the $\binom{t+r-1}{r-1}$ distributions given in Lemma~\ref{lm:Stype_FJFA}.
  In FJ-FSM, all requests are served with the fastest possible service time.
  Response time of FJ-FSM therefore serves as a lower bound for $T_{\text{FJ-FA}}$.
  As the FJ-FSM model implements an M/M/1 queue with arrival rate $\lambda$ and service time distribution $\Exp{(t+1)\mu}$, we get the result.
\end{proof}
\noindent
The immediate corollary of Lemma~\ref{lm:T_FJFA_lb_wFSM} gives a lower bound on the average download time in FJ-FA as $\E{T_{\text{FJ-FA}}} \geq 1/\left((t+1)\mu - \lambda\right)$ for $\lambda < (t+1)\mu$.

\section{Proofs of Theorems~\ref{thm:T_FJGA_ulb} and~\ref{thm:ET_FJGA_lub}}
\label{sec:proof-main-theorems}
\begin{proof}[Proof of Theorem~\ref{thm:T_FJGA_ulb}]
  \noindent
  \textbf{Lower bounds $(a)$ and $(b)$}:
  Consider an enlarged system that consists of $k$ copies, system-$i$ for $i = 1, \dots, k$, of the original FJ system.
  Let us forward the requests for object $i$ only to system-$i$.
  Then, system-$i$ is equivalent to a FJ-FA operating under the arrival rate of $p_i \lambda$.
  Each request will experience a smaller response time in this enlarged system than in the original. Hence the lower bound $(a)$.
  The Fast-Split-Merge model gives a lower bound on the response time of system-$i$ (Lemma~\ref{lm:T_FJFA_lb_wFSM}), which together with $(a)$ gives $(b)$.
  \textbf {Upper bounds $(c)$ and $(d)$} follow from Lemma~\ref{lm:T_FJ_ub_wSM}.
\end{proof}

\begin{proof}[Proof of Theorem~\ref{thm:ET_FJGA_lub}]
  Lower bound $(a)$ given on $T_{\text{FJ-GA}}$ in \eqref{eq:T_FJGA_lb} is a mixture distribution with components distributed as $\Exp{(t+1)\mu - p_i \lambda}$ for $i = 1, \dots, k$.
  Expected value of this mixture distribution is the lower bound for $\E{T_{\text{FJ-GA}}}$.
  By $(d)$ given in \eqref{eq:T_FJGA_ub}, average response time $\E{T_{\text{FJ-SM}}}$ in FJ-SM system yields an upper bound on $\E{T_{\text{FJ-GA}}}$.
  FJ-SM implements an M/G/1 queue with the service time distribution $S_{\text{slowest}}$ (Section~\ref{sec:FJSM}).
  Then the Pollaczek-Khinichin formula~\cite{Tijms} gives 
  \begin{equation}
    \E{T_{\text{FJ-SM}}} = \E{S_{\text{slowest}}} + \frac{\lambda \E{S_{\text{slowest}}^2}}{2\Bigl(1 - \lambda \E{S_{\text{slowest}}}\Bigr)}.
  \label{eq:ET_SM}
  \end{equation}
  $\E{S_{\text{slowest}}}$ is given in Theorem~\ref{thm:T_avail_lowtraff}.
  We find $\E{S_{\text{slowest}}^2}$ as
  \begin{longaligned}[\label{eq:ES02}]
    \E{S_{\text{slowest}}^2}& \stackrel{(i)}{=} \int_{0}^{\infty} 2s \;\Prob{S_{\text{slowest}} > s} \dx{s}\longalignedtag \\
    &\stackrel{(ii)}{=} \int_{0}^{\infty} 2s \exp(-\mu s) \left(1-\left(1 - \exp(-\mu s)\right)^r \right)^t \dx{s} \\
    &\stackrel{(iii)}{=} \sum_{j=0}^{t}\binom{t}{j} (-1)^j \sum_{l=0}^{r j} (-1)^l \binom{r j}{l} \frac{2}{\mu^2(l+1)^2},
  \end{longaligned}
  where (i) holds because $\E{X^2} =  \int_{0}^{\infty} 2s \;\Prob{X > s} \dx{s}$ for a non-negative r.v. $X$,
  (ii) comes from \eqref{eq:S_fastest_slowest}, and
  (iii) follows from the binomial expansion of $\left(1-\left(1 - \exp(-\mu s)\right)^r \right)^t$ and interchanging the order of integration and summation.
  Finally, substituting in \eqref{eq:ET_SM} the first and second moments of $S_{\text{slowest}}$ gives us the upper bound in \eqref{eq:ET_FJGA_lub}.
\end{proof}

\begin{proof}[Proof of Lemma~\ref{lm:T_FJGA_leq_T_FJFA}]
  Consider a sample arrival sequence $L$ of
  requests in the FJ-GA system.
  Each request-$i$ denotes a pair $(t_i, o_i)$ where $t_i$ is the arrival time and $o_i$ is the requested object.
 Let $\ell$ be the index of the first request in $L$ that is asking for an object different from the first requested object $o_1$. Let $\Gamma$ be an operator such that $\Gamma(L)$ returns a copy of $L$ with only one modification: the object being asked by request-$\ell$ is changed to the first requested object $o_1$.
  Note that $\Gamma$ keeps the request arrival times $t_i$ the same.
  Let $L^{(m)}$ be the sequence obtained by applying $\Gamma$ on $L$ repeatedly $m$ times. 
  
  Let $T_{L}$ be the download time under $L$ and $T_{L^{(1)}}$ under $L^{(1)}$.
  System under $L^{(1)}$ behaves as FJ-FA for requests-($\leq \ell$), that is, requests-($\leq \ell$) depart the system in order.
  However, under $L$, request-$\ell$ might finish early at the leading servers. This implies that request-$\ell$ might depart before even reaching HoL.
  Conditioned on the event that request-$\ell$ does not have an early departure, it is not difficult to show that $T_{L^{(1)}}$ is stochastically the same as $T_{L}$ when code locality is two.
  Further, conditioned on the event that that request-$\ell$ has an early departure, it is not difficult to show that $T_{L} \leq T_{L^{(1)}}$ stochastically when code locality is two. We skip showing these two observations here due to space constraints.
  This overall gives us $T_{L} \leq T_{L^{(1)}}$.
  
  Using the same arguments given in the previous paragraph, 
  {we can show that} $T_{L^{(i)}} \leq T_{L^{(i+1)}}$ for all $i$.
  Let $L^{(*)}$ denote a sequence after applying $\Gamma$ on $L$ sufficient number of times such that $L^{(*)}$ has all requests asking for the same object.
  Notice that $L^{(*)}$ is a sample arrival sequence for FJ-FA.
  The previous inequality we showed implies $T_{L} \leq T_{L^{(1)}} \leq T_{L^{(2)}} \leq \ldots \leq T_{L^{(*)}}$.
  Let us also define $\mathbb{L}$ (resp. $\mathbb{L}^{(*)}$) as the set of all request arrival sequences for FJ-GA (resp. FJ-FA).
  Every $L$ in $\mathbb{L}$ will be transformed into an $L^{(*)}$ in $\mathbb{L}^{(*)}$ by repeated application of operator $\Gamma$. Thus, $\Gamma$ is a surjective function from $\mathbb{L}$ to $\mathbb{L}^{(*)}$.
  Let us define the subset of all $L$ in $\mathbb{L}$ that map to $L^{(*)}$ as $\Gamma^{-1}(L^{(*)})$.
  Then it is easy to see
  \begin{equation}
    \Prob{L^{(*)}} = \sum_{L \in \Gamma^{-1}(L^{(*)})} \Prob{L}.
  \label{eq:PrLstar_eq_PrGammaInvLstar}
  \end{equation}
  where $\Prob{L}$ (resp. $\Prob{L^{(*)}}$) denotes the probability of sampling $L$ (resp. $L^{(*)}$) from $\mathbb{L}$ (resp. $\mathbb{L}^{(*)}$).
  
 The FJ-GA download time distribution can be written as
\begin{longaligned}
   \label{eq:PrT_FJ_GA_UB} 
  \Prob{T_{\operatorname{FJ-GA}} > t}
  &\stackrel{(a)}{=} \sum_{L \in \mathbb{L}} \Prob{L} \Prob{T_{L}>t}\longalignedtag \\
  &\stackrel{(b)}{=} \sum_{L^{(*)} \in \mathbb{L^{(*)}}}\sum_{L \in \Gamma^{-1}(L^{(*)})} \Prob{L} \Prob{T_{L}>t} \\
  &\stackrel{(c)}{\leq} \sum_{L^{(*)} \in \mathbb{L^{(*)}}}\sum_{L \in \Gamma^{-1}(L^{(*)})} \Prob{L}  \Prob{T_{L^{(*)}}>t},
\end{longaligned} 
  where $(a)$ follows from the law of total probability,
  $(b)$ follows from the discussion previously given on $\Gamma^{-1}(L^{(*)})$, and $(c)$ follows by using $T_{L} \leq T_{L^{(*)}}$
  Then, we find
  \begin{equation*}
  \begin{split}
    \Prob{T_{\operatorname{FJ-GA}} > t}
    &\stackrel{(a)}{\leq}
    \sum_{L^{(*)} \in \mathbb{L^{(*)}}}
    \Prob{T_{L^{(*)}}>t} \sum_{L \in \Gamma^{-1}(L^{(*)})} \Prob{L}\\
    &\stackrel{(b)}{=}
    \sum_{L^{(*)} \in \mathbb{L^{(*)}}} \Prob{L^{(*)}} \Prob{T_{L^{(*)}}>t},
  \end{split}
  \end{equation*}
  where 
  $(a)$ follows from~\eqref{eq:PrT_FJ_GA_UB}
  and $(b)$ follows from \eqref{eq:PrLstar_eq_PrGammaInvLstar}.
  The proof concludes by observing that the right hand side $\sum_{L^{(*)} \in \mathbb{L^{(*)}}} \Prob{L^{(*)}} \Prob{T_{L^{(*)}}>t} = \Prob{T_{\operatorname{FJ-FA}} > t}$.
\end{proof}

\section{M/G/1 Queue Approximation for FJ-FA}
\label{sec:mg1-approximations}

\subsection{Background}
While discussing the FJ-FA system, we re-defined the request service start times (see in Section~\ref{sec:FJFA}, Definition~\ref{def:FJFA_reqservstart}). That allowed us to dissect the system dynamics and lead us to make two important observations:
i) requests depart the FJ-FA system in the order they arrive (Observation~\ref{obv:FJFA_dep_order}), and
ii) there are $\binom{t+r-1}{r-1}$ possible request service time distributions, as given in Lemma~\ref{lm:Stype_FJFA}. These two observations, together with a the one we make below, allow us to argue that the FJ-FA system can be approximated as an M/G/1 queue.

The service time distribution for a request is dictated by the system state at its service start time.
Queue lengths carry memory between the service starts of the subsequent requests. For instance, as request-$i$ starts service, if the difference between the queue lengths across all the servers is at least $2$, then this difference will be at least $1$ as request-$(i+1)$ starts service.
Therefore in general, service times are not independent across the requests.
However, very importantly, request service times are only \textit{loosely coupled}.
Once a request is finished, its outstanding redundant copies get removed from service. This helps the slow servers to catch up with the leading servers. It is ``hard'' for the leading servers to keep leading as they compete with every other server.
Queues across all the servers are therefore expected to frequently level up.
Every time the queues level up while they are non-empty corresponds to an epoch at which a request start service (i.e., moving to HoL) with type-$0$ service time distribution.
Given a time epoch $t$ at which the queues level up, requests that move to HoL before $t$ or those that move after $t$ have independent service time distributions.
Therefore, queue levelling time epochs break the dependence between the service times.
Given that such time epochs occur frequently, request service times constitute a series of independent small-size batches.
\begin{observation}
  The FJ-FA system experiences frequent time epochs across which request service times are independent.
\label{obv:FJFA_freq_renewals}
\end{observation}

The observations we have made so far lead to \emph{an approximate method} for analyzing the FJ-FA system.
Requests depart the system in the order they arrive (Observation~\ref{obv:FJFA_dep_order}), hence the system as a whole acts as a FCFS queue.
There are $\binom{t+r-1}{r-1}$ possible distributions for the request service times (Lemma~\ref{lm:Stype_FJFA}). Although request service times are not independent, they are loosely coupled (Observation~\ref{obv:FJFA_freq_renewals}). Putting all these together, we propose the following approximation for the  FJ-FA system. 

\begin{approximation}
The FJ-FA system can be approximated as an M/G/1 queue, and it holds that
  \begin{equation}
    \E{T_{\text{FJ-FA}}} \approx \E{T_{\text{M/G/1}}} = \E{S} + \frac{\lambda \E{S^2}}{2(1 - \lambda \E{S})}.
  \label{eq:ET_FJFA_approx_wPK}
  \end{equation}
  Here, the moments of $S$ are given as
  \begin{equation}
  \begin{split}
	\E{S} &= \sum_{\bm{\nu} \in \mathbb{N}} \Prob{\operatorname{Type-}\bm{\nu}\operatorname{ service}} \E{S_{\bm{\nu}}}, \\
	\E{S^2} &= \sum_{\bm{\nu} \in \mathbb{N}} \Prob{\operatorname{Type-}\bm{\nu}\operatorname{ service}} \E{S^2_{\bm{\nu}}},
  \end{split}
  \label{eq:FJFA_servmoments}
  \end{equation}
  where the set $\mathbb{N}$ is defined in \eqref{eq:set_nu}, 
  $\Prob{\operatorname{Type-}\bm{\nu}\operatorname{ service}}$ is the probability that the service time of an arbitrary request is sampled from $S_{\bm{\nu}}$, whose distribution is in Lemma~\ref{lm:Stype_FJFA}.
\label{approx:FJFA_mg1}
\end{approximation}

Note that \eqref{eq:ET_FJFA_approx_wPK} follows from the Pollaczek-Khinichin formula~\cite{Tijms}, and \eqref{eq:FJFA_servmoments} follows from Lemma~\ref{lm:Stype_FJFA}. 
Moreover, the approximation becomes exact for the system with locality one, that is, when the Fork-Join content access model is implemented on the system with replicated storage \cite{CodingForFastContentDownload:JoshiLS12}.

In the M/G/1 approximation given above, the only unknown quantity is $\Prob{\operatorname{Type-}\bm{\nu}\operatorname{ service}}$.
Recall that we study the system dynamics in steady state. Given that, $\Prob{\operatorname{Type-}\bm{\nu}\operatorname{ service}}$ denotes the probability of serving request $i$ with type-$\bm{\nu}$ distribution in the limit $i \to \infty$.
By ergodicity the limiting value of this probability is equal to the limiting fraction $f_{\bm{\nu}}$ of the requests served with type-$\bm{\nu}$ distribution \cite{PASTA:Wolff82}.
We rely on $f_{\bm{\nu}}$ in the derivations presented in the sequel.

\subsection{FJ-FA with Locality Two}
\label{subsec:mg1_r2}
We next focus on the FJ-FA model for availability codes with locality two. This class of codes is of interest because they are minimally different from replication with locality one.

Given that recovery groups consist of two servers, there can be at most one leading server in each.
Hence a request can have at most one early departure at each of the recovery groups. (See Section~\ref{sec:FJFA} for the definition of early departing request copies.)
Request service type vector $\bm{\nu}$ in this case is given by $(\nu_0, \nu_1)$ where $\nu_0$ (resp. $\nu_1$) denotes the number of recovery groups at which the request has no (resp. one) early departing copy.
Given that $\nu_0 + \nu_1$ is fixed and equal to $t$, it is sufficient to only keep track of $\nu_1$ to determine the service time distribution of a request.
In other words, service time distribution for a request is defined by its number of early departing copies, \ie, number of recovery groups at which the request had one early departure.
Given that $\nu_1$ lies in $[0, t]$, the set $\mathbb{S}$ of all possible service time distributions is of size $t+1$.
If a request has an early departing copy at $i$ recovery groups before it moves to HoL, i.e., $\nu_1 = i$, then the request will be served with type-$i$ service time distribution. (Here, we follow the convention introduced on service types in Section~\ref{sec:FJFA}.)
Type-$i$ service time distribution is given by
\begin{equation}
  \Prob{S_i > s} = \exp(-\mu s)^{i+1} \Bigl(1 - (1 - \exp(-\mu s))^2\Bigr)^{t-i}
\label{eq:Pr_Si_g_s}
\end{equation}
for $i = 0, 1, \dots, t$.
We have $\Prob{S_{i+1} > s}/\Prob{S_i > s} < 1$ for $s > 0$.
This implies that $S_i$ are stochastically ordered as $S_0 > S_1 > \dots > S_t.$
Recall from Section~\ref{sec:FJFA} that for the code with general locality,
we have a partial ordering between the service time distributions.
This partial ordering turns into a complete ordering in this case with locality of two.

Deriving the request service time probabilities $f_i$ requires an exact analysis of the system, which is intractable due to the state explosion problem.
We next conjecture a relation between $f_i$ for the system with locality two. This relation will be used later to derive several estimates for $f_i$.
\begin{conjecture}
  In the FJ-FA system with locality two, $f_{i-1} > f_i$ for $i = 1, \dots, t$.
\label{conj:FJFA_r2_fjs}
\end{conjecture}

\begin{figure}[ht]
  \centering
  \begin{subfigure}
    \centering
    \includegraphics[width=0.9\textwidth, keepaspectratio=true]{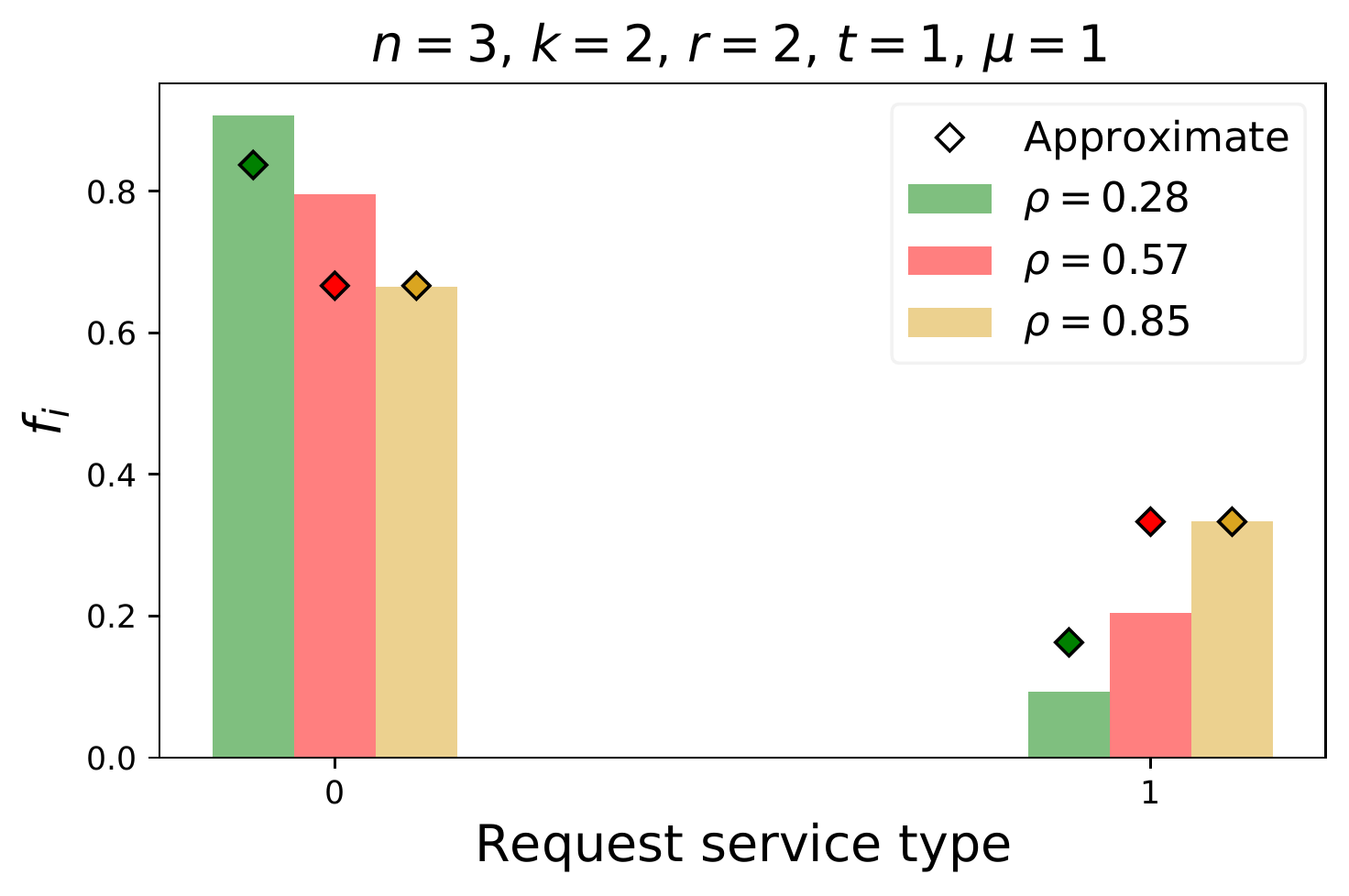}
  \end{subfigure}
  \begin{subfigure}
    \centering
    \includegraphics[width=0.9\textwidth, keepaspectratio=true]{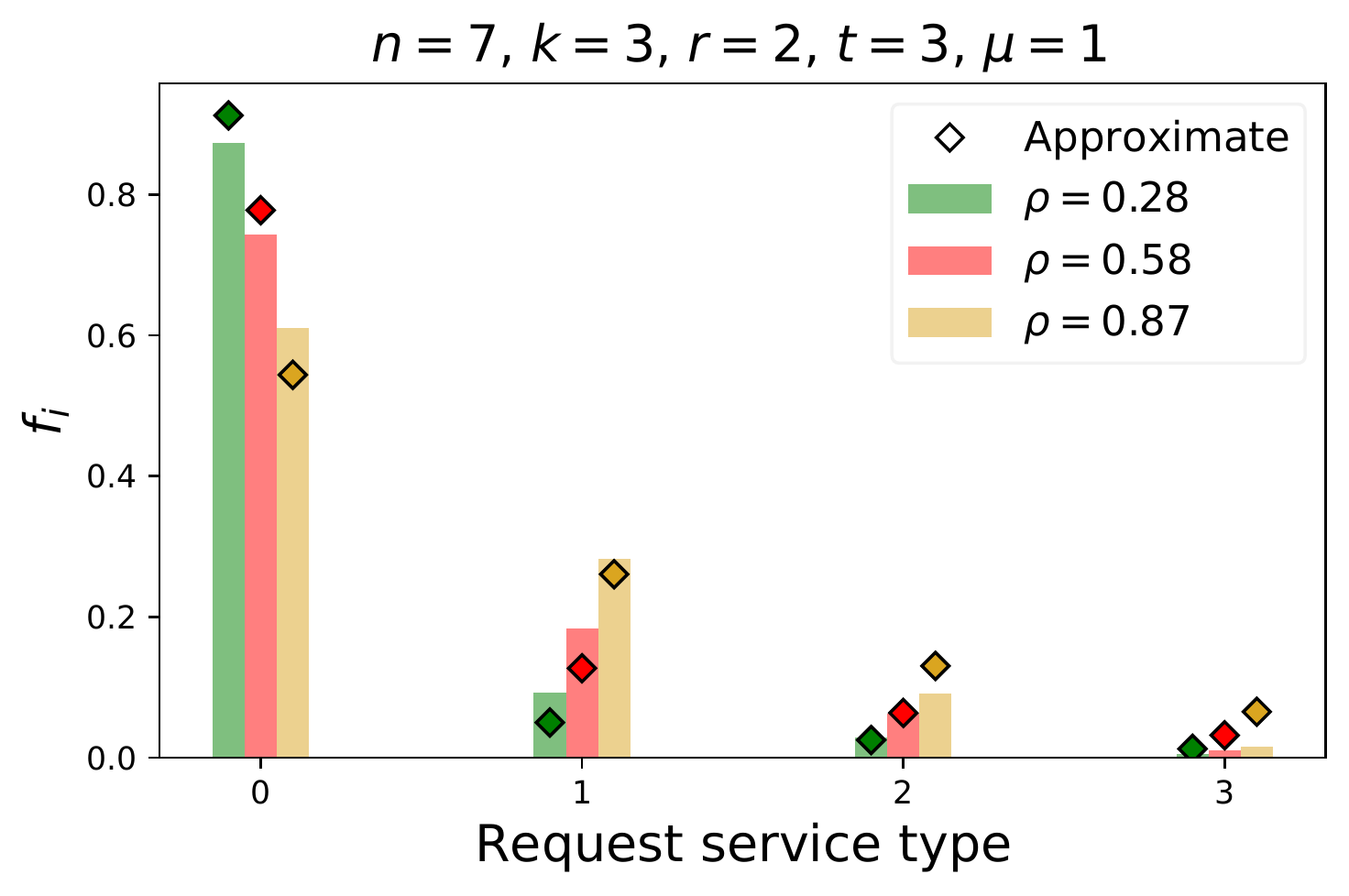}
  \end{subfigure}
  \caption{Simulated request service time probabilities $f_i$ for systems with (3, 2, 2, 1)- and (7, 3, 2, 3)-availability codes. Here $\rho$ is the average system load.}
\label{fig:conjecture}
\vspace{-4mm}
\end{figure}

We next briefly discuss the reasoning behind the conjecture.
Observation~\ref{obv:FJFA_freq_renewals} states that the queues at the servers frequently level up. This is because the leading servers in the recovery groups compete with every other server in the system to keep leading. For a request to have type-$i$ distribution for service, it needs to have one early departure at exactly $i$ recovery groups.
This requires one server in each of the $i$ recovery groups to be leading, which gets less likely for larger $i$.
We have validated the conjecture with extensive simulations. Fig.~\ref{fig:conjecture} shows the simulated values for $f_i$ in a system that employs the binary Simplex code with availability one or three, and locality two.
Furthermore, we found a strong pointer for the conjecture (can be found in \cite{PhDThesis:Aktas20}). This pointer says that given a request is served with type-$i$ distribution, the subsequent request is more likely to be served with type-$j$ distribution for $j < i$.
We prove the conjecture for the system with availability one and locality two. This is implied by the bounds given on $f_i$ in Theorem~\ref{thm:FJFA_r2_t1_fi_bound}.

Fig.~\ref{fig:conjecture} shows that, as the offered load increases, the frequency of type-$i$ service time increases for larger $i$, which
agrees with the discussion above. To have type-$i$ distribution, a request needs to have one early departure at $i$ recovery groups. This is possible only if there are $i$ leading servers. Recovery servers can go ahead of their peers only if their queues are non-empty. The queues are more likely to build up under greater offered load. This allows the leading servers to progress even further than they could under smaller offered load.
Using Conjecture \ref{conj:FJFA_r2_fjs}, we find estimates for $f_i$.
( see\cite{PhDThesis:Aktas20} for details). Then substituting these estimates in the M/G/1 approximation (cf.~\eqref{eq:ET_FJFA_approx_wPK}), we derive the following approximation.
\begin{approximation}
  The FJ-FA system with locality two is approximately an M/G/1 queue with service time distribution
  \[ \Prob{S > s} =  \Bigl(1 + \sum_{i=1}^t \prod_{j=0}^{i-1} \rho_j\Bigr)^{-1} \sum_{i = 0}^t \Prob{S_i > s} \prod_{j=0}^{i-1} \rho_j, \]
  where $\rho_j$ are recursively computed as 
   ($i = 1, \dots, t$)
  \begin{equation*}
  \begin{split}
     \rho_0 &= \frac{\lambda \E{V}}{t \bigl(1 - \lambda \E{V}\bigr)}, \\
     \rho_i &= \frac{1 - \Bigl(1 - \lambda \E{V}\bigr)\bigl(1 + \sum_{k=0}^{i-1} \prod_{l=0}^{k} \rho_l \Bigr)}{\bigl(1 - \lambda \E{V}\bigr)(t-i)\prod_{k=0}^{i-1} \rho_k},
  \end{split}
  \end{equation*}
  where $\E{V} = \frac{1}{t+1}\sum_{i=0}^t \E{S_i}$ and $S_i$ are defined in \eqref{eq:Pr_Si_g_s}.
\label{approx:FJFA_r2_best_approx}
\end{approximation}

\begin{figure}[t!]
  \includegraphics[width=0.9\textwidth]{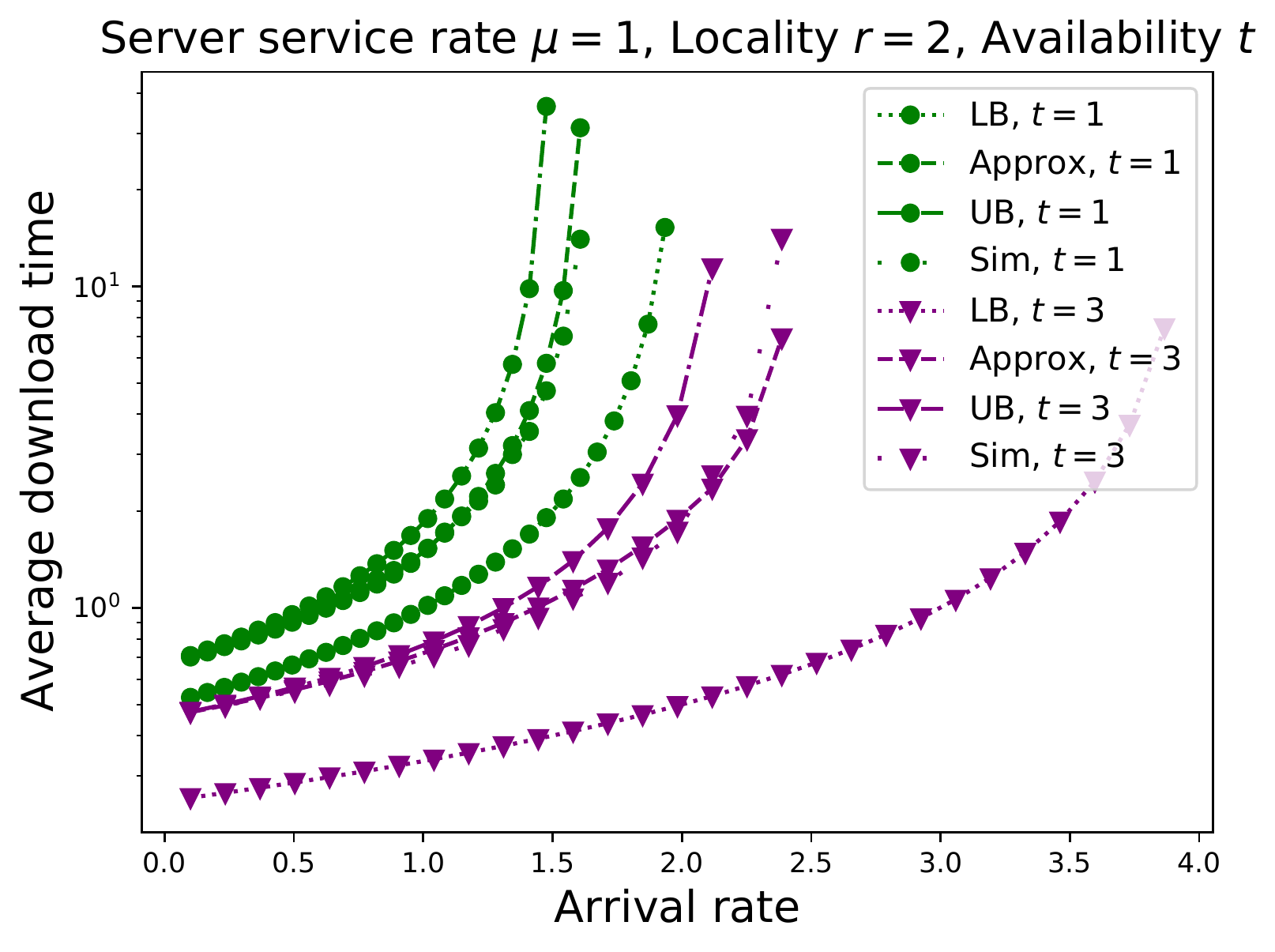}
  \caption{
  Average download time versus $\lambda$ for FJ-FA with locality two.
  \emph{Sim} refers to the simulated values.
  \emph{UB} (resp. \emph{LB}) refers to the upper (resp. lower) bounds in Lemma~\ref{lm:T_FJ_ub_wSM} (resp. \ref{lm:T_FJFA_lb_wFSM}). \emph{Approx} refers to Approximation~\ref{approx:FJFA_r2_best_approx}.}
\label{fig:plot_ET_FJFA_r2}
\vspace{-4mm}
\end{figure}

\vspace{1ex}
\noindent
\textbf{Comparison of the approximations and bounds:}
Fig.~\ref{fig:plot_ET_FJFA_r2} gives a comparison between the M/G/1 approximation (Approximation~\ref{approx:FJFA_r2_best_approx}) and the bounds presented in Section~\ref{sec:FJFA}.

\subsection{FJ-FA with Locality Two and Availability One}
\label{subsec:FJFA_r2_t1}
We here focus on availability codes with locality two and availability one.
This represents the simplest possible availability code, i.e., objects $a$, $b$ are stored together with $a+b$ over three servers. The corresponding
FJ system is thus the simplest of all FJ systems with availability codes.
Requests in this case are served with either type-$0$ or type-$1$ service time distribution.
Although the exact analysis is still formidable, the system state is simple enough to find tighter bounds on the request service time probabilities $f_0$ and $f_1$.
Substituting these into Approximation~\ref{approx:FJFA_r2_best_approx} leads to a tighter M/G/1 approximation.

The simplicity of the setting here allows us to address a more general service time model than identical $\Exp{\mu}$ r.v.'s. 
We continue assuming the service times to be independent across different servers and request copies.
However this time, we let the service time at the systematic server to be distributed as $\Exp{\gamma}$, and the service times at the recovery servers to be distributed as $\Exp{\alpha}$ and $\Exp{\beta}$.
In the following we refer to server with service rate $x$ as server-$x$.

The system state at time $\tau$ can be described with a triple $\bm{s}(\tau) = (N(\tau), (n_{\alpha}(\tau), n_{\beta}(\tau)))$.
$N(\tau)$ denotes the total number of requests in the system at time $\tau$. This is given by the number of request copies present in the systematic server.
$n_{\alpha}(\tau)$ (resp. $n_{\beta}(\tau)$) denotes by how many request copies server-$\alpha$ (resp. server-$\beta$) is leading the other recovery server. In other words, denoting the queue length at server-$x$ at time $\tau$ with len-$x(\tau)$, we can express $n_{\alpha}(\tau)$ and $n_{\beta}(\tau)$ as
\begin{equation}
\begin{split}
  n_{\alpha}(\tau) &= \max\{\text{len-}\alpha(\tau)-\text{len-}\beta(\tau), 0\}, \\
  n_{\beta}(\tau) &= \max\{\text{len-}\beta(\tau)-\text{len-}\alpha(\tau), 0\}.
\end{split}
\label{eq:n_tau}
\end{equation}

\begin{figure}[t]
  \includegraphics[width=0.9\textwidth]{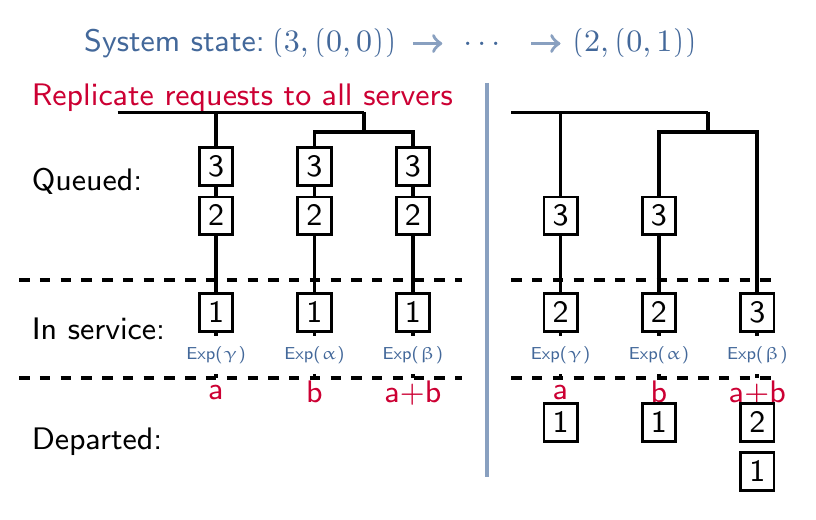}
  \caption{FJ-FA with availability one and locality two. Snapshot on the left (resp. right) illustrates the time epoch at which request $1$ (resp. $2$) starts service and its service time distribution $S_0$ (resp. $S_1$).}
\label{fig:FJFA_r2_t1_sys_snapshot}
\end{figure}

It follows that $n_{\alpha}(\tau) n_{\beta}(\tau) = 0$ for all $\tau$.
This is because there can be only one leading recovery server at any time since the system has one recovery group of two servers (availability one, locality two). (Recall the definition of leading server from the second paragraph of Section~\ref{sec:FJFA}.)
The system state is illustrated in Fig.~\ref{fig:FJFA_r2_t1_sys_snapshot} with two different snapshots of the system.

\begin{figure}[ht]
 \begin{tikzpicture}
 \node[above] at (0,0) {\includegraphics[scale=0.75]{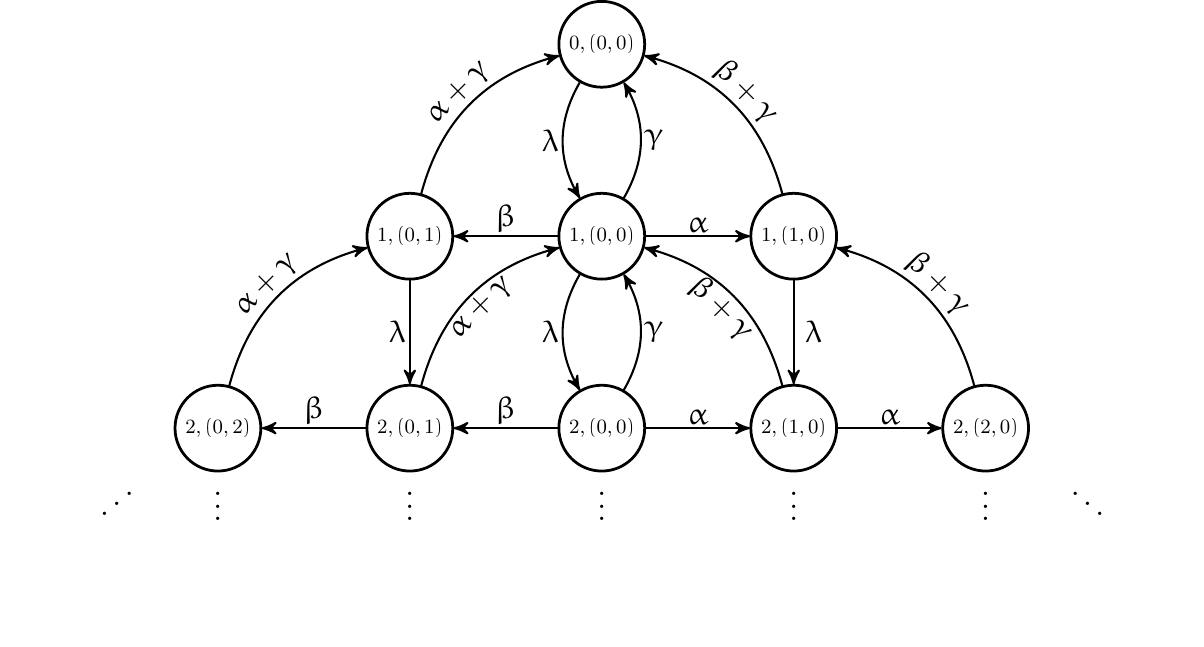}};
 \node at (0,0.2) {\includegraphics[scale=0.75]{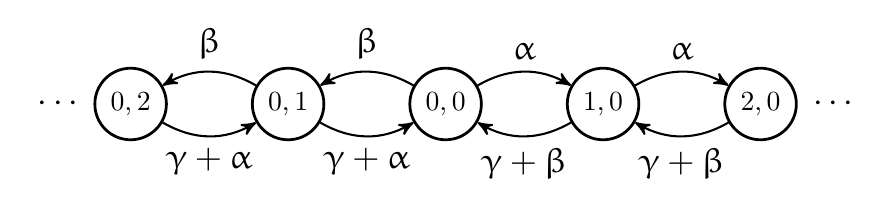}};
 \end{tikzpicture}
 \caption{Markov state process for the FJ-FA system with availability one and locality two (top), and its high traffic approximation (bottom).}
\label{fig:FJFA_r2_t1_state}
\end{figure}

The system state $\bm{s}(\tau)$ at time $\tau$ is a Markov process as illustrated in Fig.~\ref{fig:FJFA_r2_t1_state} (top).
Let us refer to $\Prob{\bm{s}(\tau) = (k, i, j)}$ by $p_{k,i,j}(\tau)$. Suppose that the system stability is imposed and $\lim_{\tau \to \infty} p_{k,i,j}(\tau) = p_{k, i, j}$. Balance equations for the system are summarized for $k,i,j \geq 0$ as
\begin{equation*}
\begin{split}
  \bigl(\gamma & \mathbbm{1}(k \geq 1) + \alpha \mathbbm{1}(i \geq 1) + \beta \mathbbm{1}(j \geq 1)\bigr) p_{k,i,j} \\
  &= \lambda \mathbbm{1}(k \geq 1, i \geq 1, j \geq 1) p_{k-1,i-1,j-1} + \gamma\; p_{k+1,i+1,j+1} \\
  &\quad + (\gamma + \alpha) p_{k+1,i+1,j} + (\gamma + \beta) p_{k+1,i,j+1}.
\end{split}
\end{equation*}
These balance equations do not permit exact analysis of the system's steady state behavior, because the state space is infinite in two dimensions.
We use two methods to analyse the process approximately: 
1) the \emph{local balance method} with a guess-based analysis \cite[Chapter 17]{QueueingBook:Harchol13}, and
2) the \textit{matrix analytic method} \cite[Chapter 21]{QueueingBook:Harchol13}, which involves truncating the process and numerically finding $p_{k, i, j}$ with an iterative procedure. The analysis is given in \cite{PhDThesis:Aktas20}.
The matrix analytic method gives the following upper bound on the average download time.
This bound is provably tighter than the Split-Merge upper bound previously given in Theorem~\ref{thm:ET_FJGA_lub}. This is because the Split-Merge model truncates the pyramid process and keeps only the central column, while in our application of the matrix analytic method, we keep the five central columns.

\begin{theorem} 
  In the FJ-FA system with availability one and locality two, the average data download time is bounded as
  \begin{equation*}
  \begin{split}
    \E{T_{\text{FJ-FA}}} &< \frac{1}{\lambda} \Bigl(\bm{\pi}_0 \bm{1}_0^T - \pi_{0,(0,0)} \\
    &\qquad\quad + \bm{\pi}_1 \left((\bm{I}-\bm{R})^{-2} + (\bm{I}-\bm{R})^{-1}\right) \bm{1}_1^T \Bigr).
  \end{split}
  \end{equation*}
  Here $\bm{1}_0 = [1, 1, 1, 1]$, $\bm{1}_1 = [1, 1, 1, 1, 1]$.
  and $\bm{1}_0^T$ and $\bm{1}_1^T$ refer to their transpose.
  Vectors $\bm{\pi}_0$ and $\bm{\pi}_i$ are given as
  \begin{equation*}
  \begin{split}
      \bm{\pi}_0 &= [\pi_{0,(0,0)}, ~\pi_{1,(0,1)}, ~\pi_{1,(0,0)}, ~\pi_{1,(1,0)}], \\
      \bm{\pi}_1 &= [\pi_{2,(0,2)}, ~\pi_{2,(0,1)}, ~\pi_{2,(0,0)}, ~\pi_{2,(1,0)}, ~\pi_{2,(2,0)}],
  \end{split}
  \end{equation*}
  where $\pi_{n_{\gamma},(n_{\alpha},n_{\beta})}$ is the steady-state probability that queue lengths at the servers are $\text{len-}\gamma = n_{\gamma}$, $\text{len-}\alpha = n_{\alpha}$ and $\text{len-}\beta = n_{\beta}$.
  Matrices $\bm{I}$ and $\bm{R}$ are $5 \times 5$. $\bm{I}$ is the identity matrix. $\bm{R}$ is numerically computed in terms of the arrival rate $\lambda$ and service rates $\gamma$, $\alpha$ and $\beta$. The algorithm is described in \cite{PhDThesis:Aktas20}.
\label{thm:FJFA_r2_t1_matrixanalytic_ub}
\end{theorem}

\vspace{0.5ex}
\noindent\textbf{High-traffic approximation:} 
We next present a method to approximately analyze the system.
As in Section~\ref{sec:FJFA}, we first find estimates for the request service time probabilities $f_0$ and $f_1$, and then substitute these in the M/G/1 approximation (cf.~\eqref{eq:ET_FJFA_approx_wPK}).
Here we are able to find tighter estimates for $f_0$ and $f_1$ by exploiting the simplicity of the system state.

Suppose that the system operates close to its stability limit, such that the servers are always busy serving a request copy. This reduces the system complexity, as we can now describe the system state keeping track of $n(\tau) = (n_{\alpha}(\tau), n_{\beta}(\tau))$ defined in \eqref{eq:n_tau}.
System state in this case implements a birth-death Markov process as shown in Fig.~\ref{fig:FJFA_r2_t1_state} (bottom).
Referring to $\lim_{\tau \to \infty} \Prob{n(\tau) = (i, j)}$ as $p_{i,j}$, and using the balance equations $\alpha \;p_{i,0} = (\gamma + \beta) p_{i+1,0}$ and $\beta \;p_{0,i} = (\gamma + \alpha) p_{0,i+1}$ for $i \geq 0$, we find the limiting state probabilities as 
\begin{align*}
  p_{0,0} &= \frac{\gamma^2-(\alpha-\beta)^2}{\gamma(\alpha+\beta+\gamma)}, ~~~~
  p_{i,0} = \Bigl(\frac{\alpha}{\beta + \gamma}\Bigr)^i p_{0,0}, \\
  p_{0,i} &= \Bigl(\frac{\beta}{\alpha + \gamma}\Bigr)^i p_{0,0} \quad\text{ for } i \geq 1.
\end{align*}

We next use these expressions to find bounds on the fraction of the requests completed by the systematic server, $w_s$, or by the recovery group, $w_r$. Recall that a request completes as soon as either its copy at the systematic server or both copies at the recovery group finish service.
For the sake of simplicity, let the service rates $\alpha$ and $\beta$ at the recovery servers be $\mu$. We keep the service rate at the systematic server fixed at $\gamma$.
\begin{theorem}
  In FJ-FA system with availability one and locality two, $w_s \geq \frac{\gamma \nu}{\gamma \nu+2\mu^2}$ and $w_r \leq \frac{\mu^2}{\gamma \nu+2\mu^2}$ for $\nu = \gamma+2\mu$.
\label{thm:FJFA_r2_t1_winning_freqs}
\end{theorem}
\begin{proof}
  Suppose the system operates close to its stability limit.
  Under this high-traffic assumption, let us refer to the values of $w_s$ and $w_r$ as $\hat{w_s}$ and $\hat{w_r}$.
  The recovery servers regularly go idle under stability. 
  Therefore, fraction of the request completions at the recovery group is smaller under high-traffic approximation than it is under stability. Hence $w_r \leq \hat{w}_r$, and $w_s = 1 - w_r$ implies $w_s \geq \hat{w}_s$.

  Next, we derive $\hat{w_s}$ and $\hat{w_r}$ from the steady state probabilities of the Markov chain embedded in the state process $n(\tau)$ (Fig.~\ref{fig:FJFA_r2_t1_state}, bottom).
  System stays at each state for an exponential duration of rate $\nu = \gamma + 2\mu$. Therefore, steady state probabilities $p_i$ of $n(\tau)$ (i.e., the limiting fraction of the time spent in state $i$) and the steady state probabilities $\pi_i$ of the embedded Markov chain (i.e., the limiting fraction of the state transitions into state $i$) are equal.
  This is easily seen by the equality
  $ \left. \pi_i = p_i \nu \middle/ \sum_{i \geq 0} p_i \nu = p_i. \right. $
  
  Let $f_s$ be the limiting fraction of the state transitions that represent request completions by the systematic server. Let $f_r$ denote the same quantity for the recovery group.
  We have
  \begin{equation*}
  \begin{split}
    f_s &= \pi_{0,0} \gamma/\nu + \sum_{i=1}^{\infty} \pi_{i,0} \;\gamma/\nu + \sum_{i=1}^{\infty} \pi_{0,i} \;\gamma/\nu = \gamma/\nu, \\
    f_r &= \sum_{i=1}^{\infty} (\pi_{0,i} + \pi_{i,0}) \mu/\nu = 2\left(\mu/\nu\right)^2.
  \end{split}
  \end{equation*}
  The limiting fraction of the state transitions that correspond to request departures is 
  $f_d = f_s + f_r = (\gamma \nu + 2\mu^2)/\nu^2$.
  Thus the fraction of requests completed by the systematic server and the recovery group are
  \[
    \hat{w}_s = \frac{f_s}{f_d} = \frac{\gamma \nu}{\gamma \nu + 2\mu^2}, \quad
    \hat{w}_r = \frac{f_r}{f_d} = \frac{2\mu^2}{\gamma \nu + 2\mu^2}.
    \qedhere
  \]
\end{proof}

\begin{figure}[t]
  \includegraphics[width=0.9\textwidth]{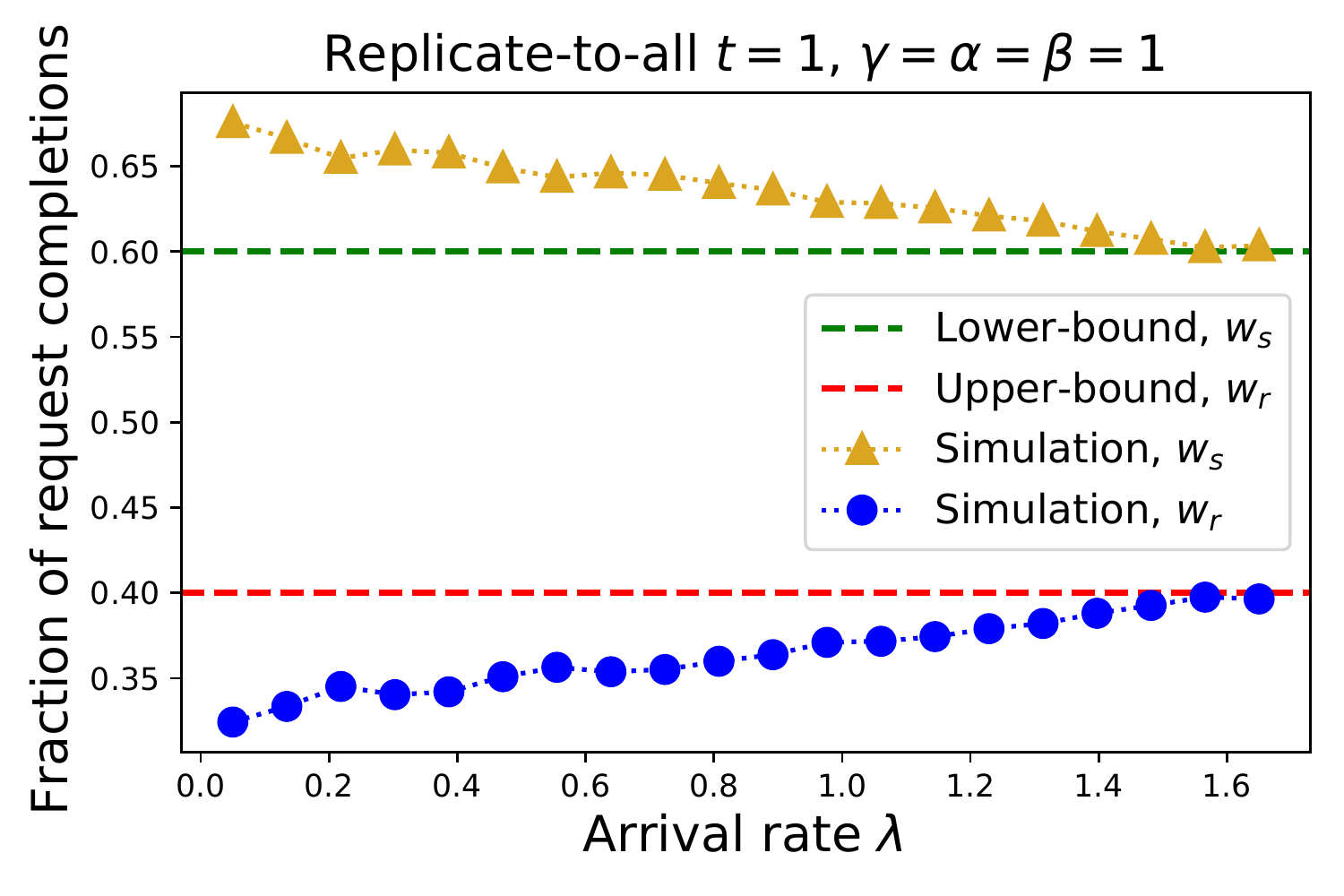}
  \caption{Simulated fraction of the request completions by the systematic server, $w_s$, and by the recovery group, $w_r$.
  The lower and the upper bound are computed using the expressions given in Theorem~\ref{thm:FJFA_r2_t1_winning_freqs}.}
\label{fig:FJFA_r2_t1_winning_freqs}
\vspace{-2ex}
\end{figure}

Fig.~\ref{fig:FJFA_r2_t1_winning_freqs} shows the general tightness of the high traffic bounds given in Theorem~\ref{thm:FJFA_r2_t1_winning_freqs}, which increases with the offered load as expected.
We next find bounds on the request service time probabilities $f_0$ and $f_1$ using the high-traffic approximation.
Note that bounds given below prove Conjecture~\ref{conj:FJFA_r2_fjs} for the system with locality two and availability one.

\begin{theorem}
  In the FJ-FA system with availability one and locality two, 
  $f_0 \geq \frac{\gamma \nu}{\gamma \nu + 2\mu^2}$ and $f_1 \leq \frac{2\mu^2}{\gamma \nu + 2\mu^2}$ for $\nu = \gamma+2\mu.$
\label{thm:FJFA_r2_t1_fi_bound}
\end{theorem}
\begin{proof}
  Suppose that the system operates close to its stability limit.
  Under this high-traffic assumption, let us refer to the values of $f_0$ and $f_1$ as $\hat{f_0}$ and $\hat{f_1}$.
  Under stability, system has to empty out regularly. Every request that finds the system empty, makes type-$0$ service start.
  However, such idle periods never happen under high-traffic approximation.
  That is why $f_0$ under the high traffic approximation is smaller than its value under stability.
  Thus we conclude $f_0 \geq \hat{f}_0$ and $f_1 \leq \hat{f}_1$.
  
  Next, we derive $\hat{f_0}$ and $\hat{f_1}$
  using the steady state probabilities of the Markov chain embedded in $n(\tau)$ (Fig.~\ref{fig:FJFA_r2_t1_state}, bottom).
  State transitions towards state $(0,0)$ represent request completions. Let $f_d$ be the fraction of such state transitions.
  Under high-traffic, every time the request at HoL departs there is a subsequent request that starts service.
  Thus, every time system transitions into $(0,0)$ (or any other transition towards it), a new request makes a type-$0$ (resp.\ type-$1$) service start.
  Let $f_{\to 0}$ and $f_{\to 1}$ be the fraction of state transitions that represent type-$0$ and type-$1$ service starts. Then
  \begin{equation*}
  \begin{split}
    f_d &= \pi_{0,0} \gamma/\nu + \sum_{i=1}^{\infty} (\pi_{0,i} + \pi_{i,0}) (\mu + \gamma)/\nu \\
    &= (2\mu^2 + 2\mu \gamma + \gamma^2)/\nu^2 \\
    f_{\to 0} &= \pi_{0,0} \gamma/\nu + \pi_{1,0} (\mu + \gamma)/\nu + \pi_{0,1} (\mu + \gamma)/\nu \\
    &= \pi_{0,0} \Bigl(\frac{\gamma}{\nu} + \frac{2\mu}{\mu + \gamma} \frac{\mu + \gamma}{\nu}\Bigr) = \pi_{0,0} = \frac{\gamma}{\gamma+2\mu}.
  \end{split}
  \end{equation*}
 Thus, the limiting fraction $\hat{f}_0$ (resp.\ $\hat{f}_1$) of the requests that make type-$0$ (type-$1$) service start are, for $\nu = \gamma + 2\mu$,\\
  \[ \hat{f}_0 = \frac{f_{\to 0}}{f_d} = \frac{\gamma \,\nu}{\gamma \,\nu + 2\mu^2}, ~~ \hat{f}_1 = 1 - \hat{f}_0 = \frac{2\mu^2}{\gamma \,\nu + 2\mu^2}
  \qedhere
  \]
\end{proof}

\begin{figure}[t]
  \includegraphics[width=0.9\textwidth]{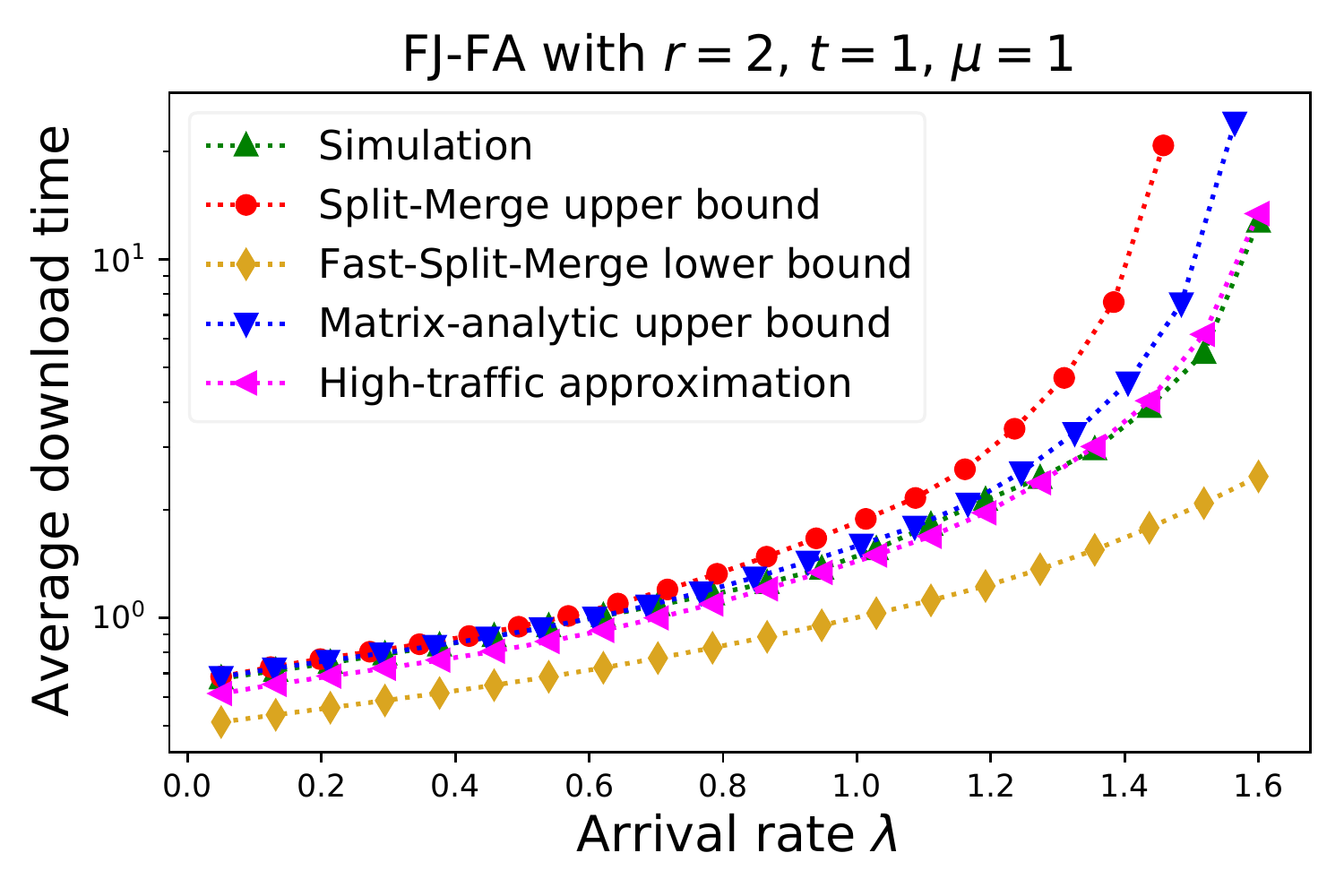}
  \caption{Comparison of the bounds on average download time versus $\lambda$ for FJ-FA with availability one and locality two.
  Split-Merge and Fast-Split-Merge bounds are given in lemmas~\ref{lm:T_FJ_ub_wSM} and \ref{lm:T_FJFA_lb_wFSM}.
  Matrix analytic bound is given in Theorem~\ref{thm:FJFA_r2_t1_matrixanalytic_ub}.
  High-traffic approximation comes from substituting in \eqref{eq:ET_FJFA_approx_wPK} the bounds given in Theorem~\ref{thm:FJFA_r2_t1_fi_bound}.}
\label{fig:ET_FJFA_r2_t1_hightraff_approx}
\vspace{-2ex}
\end{figure}

\vspace{0.5ex}
\noindent
\textbf{Comparison of approximations}:
We approximate the FJ-FA system as an M/G/1 queue (Approximation~\ref{approx:FJFA_mg1}), which together with the PK formula gives us an approximate expression for the average download time \eqref{eq:ET_FJFA_approx_wPK}.
This approximation requires the first and second moments of the service time distribution \eqref{eq:FJFA_servmoments}.
Substituting the bounds in Theorem~\ref{thm:FJFA_r2_t1_fi_bound} in place of the actual probabilities $f_0$ and $f_1$ yields the following lower bounds on the service time moments:
\begin{equation*}
\begin{split}
  \E{S} &\geq \frac{1}{3}\Bigl(\frac{2}{\gamma+\mu} - \frac{1}{\gamma+2\mu}\Bigr) + \frac{2}{3} \frac{1}{\gamma+\mu}, \\
  \E{S^2} &\geq \frac{1}{3} \Bigl(\frac{4}{(\gamma+\mu)^2} - \frac{2}{(\gamma+2\mu)^2}\Bigr) + \frac{2}{3} \frac{2}{(\gamma+\mu)^2}.
\end{split}
\end{equation*}
Substituting these bounds  in \eqref{eq:ET_FJFA_approx_wPK} gives a lower bound on the download time.
However, this lower bound can only be treated as an approximation since \eqref{eq:ET_FJFA_approx_wPK} is not exact but an approximation.
As shown in Fig.~\ref{fig:ET_FJFA_r2_t1_hightraff_approx}, this approximation performs very well in predicting the actual average download time, especially compared to the Split-Merge and Fast-Split-Merge bounds given in lemmas~\ref{lm:T_FJ_ub_wSM} and \ref{lm:T_FJFA_lb_wFSM}.

\section{Conclusions and Future Directions}

Storage systems with availability codes allow simultaneous downloads of each data object from multiple recovery groups. However, downloading an object from a recovery group requires fetching the coded objects from all the servers in the group. Such downloads are slower than those from a single server because of straggling servers. In this paper, we asked if availability codes improve hot data download performance. We adopted the Fork-Join (FJ) access scheme that employs redundant requests to mitigate stragglers' impact.

We found that availability codes lie between replication and MDS codes in the storage overhead vs. download latency tradeoff. Also, we found that availability codes achieve a favorable tradeoff between the storage overhead and download latency compared with the state-of-the-art erasure codes. We made these observations by deriving expressions for the download latency. An exact analysis was possible under the low traffic regime. Otherwise, availability codes give rise to multi-layer inter-dependent FJ queues. The state space explosion makes the exact analysis intractable in this case. Here, we derived bounds and approximations on the download time by devising systems that are tractable variants of the FJ system.
For the case of locality two, which is minimally different from replication (i.e., locality one), we conjectured an order relationship between the service time probabilities, 
and derived approximations on the download time based on this conjecture. We demonstrated, via simulations, that these approximations are close to the download time.

We considered the FJ access model that replicates every request to all the servers upon arrival. This redundancy strategy treats all the requests uniformly, resulting in fairness. However, it aggressively adds redundant load on the system, which might be suboptimal. 
An interesting future direction is to investigate strategies that employs redundancy selectively based on the system state such as the current load offered on the system.
Another future direction is to devise redundancy policies that are provably optimal for systems employing a given availability code. 
Overall, the ultimate goal is to design, for a given set of parameters, coding schemes and accompanying redundancy strategies  that are optimal in terms of the download delay. 


\balance
\bibliographystyle{IEEEtran}
\bibliography{references}

\end{document}